\documentclass[12pt,preprint]{emulateapj}  

\usepackage{graphicx}
\usepackage{txfonts}
\usepackage{hyperref}

\newcommand{\va}{v_{\mathrm{A}}}
\newcommand{\vae}{v_{\mathrm{A,e}}}
\newcommand{\vai}{v_{\mathrm{A,i}}}

\newcommand{\der}{{\rm d}}

\newcommand{\rhoi}{\rho_{\rm i}}
\newcommand{\rhoe}{\rho_{\rm e}}
\newcommand{\xii}{\mbox{\boldmath{$\xi$}}}
\newcommand{\omegaA}{\omega_{\rm A}}
\newcommand{\omegaAi}{\omega_{\rm A,i}}
\newcommand{\omegaAe}{\omega_{\rm A,e}}

\newcommand{\rhotr}{\rho_{\rm tr}}

\usepackage{natbib}

\begin{document}

	\title{Magnetohydrodynamic kink  waves  in  nonuniform solar flux tubes:\\ phase mixing and energy cascade to small scales}

 \author{Roberto Soler and Jaume Terradas}
     \affil{Departament de F\'isica, Universitat de les Illes Balears,
               E-07122, Palma de Mallorca, Spain}
              \email{roberto.soler@uib.es}

 \begin{abstract}
 Magnetohydrodynamic (MHD) kink waves are ubiquitously observed in the solar atmosphere. The propagation and damping of these waves may play relevant roles for the  transport and dissipation of energy in the solar atmospheric medium. However, in the atmospheric plasma dissipation of  transverse MHD wave energy by viscosity or  resistivity needs very small spatial scales to be efficient. Here, we theoretically investigate the generation of small scales in nonuniform solar magnetic flux tubes due to phase mixing of   MHD kink waves. We go beyond the usual approach based on the existence of a   global quasi-mode that is damped in time due to resonant absorption. Instead, we use a  modal expansion to express the  MHD kink wave as a superposition of  Alfv\'en continuum modes that are phase mixed as time evolves. The comparison of the two techniques evidences that the  modal analysis is more physically transparent and describes both the damping of  global kink motions and the building up of small scales due to phase mixing.  In addition, we discuss that the processes of resonant absorption and phase mixing are intimately linked. They represent two aspects of the same underlying physical mechanism: the energy cascade from large scales to small scales due to naturally occurring plasma and/or magnetic field inhomogeneities. This process may provide the necessary scenario for  efficient dissipation of transverse MHD wave energy in the solar atmospheric plasma.
  \end{abstract}



\section{INTRODUCTION}

Recent high-resolution observations indicate that transverse magnetohydrodynamic (MHD) waves are ubiquitous throughout the solar atmosphere \citep[e.g.,][]{depontieu2007,tomczyk2007,lin2009,okamoto2011,mcintosh2011,kuridze2012,depontieu2012,morton2013,morton2014}. From the theoretical point of view, the observed waves are interpreted as MHD kink waves on magnetic flux tubes \citep[see, e.g.,][]{edwin1983,goossens2012}. Both observations and theoretical aspects of these waves have been recently reviewed by \citet{demoortel2012} and \citet{mathioudakis2013}. It is believed that the propagation of transverse MHD waves and their dissipation may play relevant roles for the heating and energy transport in the solar atmospheric plasma \citep[see, e.g.][]{cargill2011,mcintosh2011,parnell2012,hahn2014}. Heating may be caused by wave energy dissipation due to Ohmic diffusion and/or shear viscosity. However, in the solar atmosphere Ohmic and viscous dissipation  are inefficient unless very short spatial scales are involved \citep[see][]{goedbloed2004}. For instance, the magnetic Reynolds number, $R_m$, indicates the efficiency of Ohmic diffusion. It can be defined as 
\begin{equation}
R_m = \frac{l_0 v_0}{\eta}, \label{eq:reynolds}
\end{equation}
where $l_0$ and $v_0$ are characteristic length and velocity scales of the plasma, respectively, and $\eta$ is the Ohmic diffusivity. We can relate $v_0$ with the plasma Alfv\'en velocity ($\sim 1000$~km~s$^{-1}$ in the solar corona) and $l_0$ with the length scale of the wave perturbations. The Ohmic diffusivity depends on the plasma properties and is extremely small for atmospheric plasma conditions \citep[see, e.g.,][]{priest2014}. This results in  $R_m \sim 10^{12}$ for typical parameters in the solar corona. However, $R_m \lesssim 1$ is needed for efficient dissipation. Such small values of the magnetic Reynolds number can only be achieved in the solar atmosphere if $l_0$ is very small.  Thus, some physical process able to transfer wave energy from large spatial scales to small spatial scales is needed for Ohmic dissipation of transverse MHD wave energy  to be efficient in the solar atmospheric plasma. The same result holds in the case of dissipation due to shear viscosity.

Resonant absorption and phase mixing are two interrelated mechanisms that may be involved in the energy cascade to small scales. Resonant absorption is an ideal process due to naturally occurring plasma and/or magnetic field inhomogeneities.  In solar atmospheric plasmas, heating involving the process of resonant absorption was first suggested by \citet{ionson1978} and has been extensively investigated in the literature afterwards. The theory of resonant MHD waves in the solar atmosphere has been reviewed by \citet{goossens2011}. Resonant absorption has a strong impact on the temporal evolution of MHD kink waves in nonuniform flux tubes because it produces a  radial flux of wave energy towards the nonuniform boundary of the flux tube \citep[see, e.g.,][]{tataronis1975,abels1979,poedts1989,arregui2011,pascoe2013,goossens2013}.   The net result  of the radial flux of energy is the simultaneous  damping of the global kink motion and the  growing of   rotational motions in the nonuniform boundary of the tube.  Although there is no direct observational evidence for this mechanism in the solar plasma, the process has been checked by time-dependent numerical simulations in which the   MHD equations are evolved in time \citep[see, e.g.,][]{terradas2006,terradas2008,pascoe2013,goossens2014}. The numerical simulations  show how the energy of the global kink motion is  transferred to the nonuniform boundary of the tube where small spatial scales are generated by phase mixing \citep[see, e.g.,][]{pritchett1978,heyvaerts1983,cally1991}.  The process of phase mixing occurs by the fact that the plasma is transversely nonuniform to the magnetic field direction, and so the Alfv\'en frequency is spatially dependent.   Alfv\'en waves propagating in adjacent magnetic surfaces become more and more out of phase as time progresses. The consequence of this mixing of phases is the continuous decrease of the wave perturbations spatial scale across the magnetic field \citep[e.g.,][]{heyvaerts1983,mann1995}. In this scenario, the process of phase mixing would hypothetically keep working until the generated spatial scales become sufficiently small for Ohmic diffusion and/or viscosity to be operative. Then, efficient plasma heating due to Ohmic/viscous dissipation of wave energy can take place \citep[see, e.g.,][]{poedts1989,poedts1990,poedts1994,ofman1995}.

The goal of this paper is to investigate the time-dependent behaviour of MHD kink waves in nonuniform solar flux tubes by focussing on the processes of phase mixing and  building up of small scales. Therefore, we do not deal with the process of plasma heating itself, but we are interested in the previous step where the energy cascade form large scales to small scales takes place. Here we use a modal expansion to analytically  express the  MHD kink wave as a superposition of discretized modes of the Alfv\'en continuum. The full time-dependent evolution of MHD kink waves in nonuniform tubes has been investigated in the past by solving the initial-value problem analytically using the Laplace transform \citep[see][]{lee1986,rudermanroberts2002} and from a purely numerical point of view \citep[see, e.g.,][among others]{terradas2006,terradas2008,pascoe2013}. However, in the present paper we choose to follow a semi-analytic process that has some advantages compared to full numerical simulations and is much simpler than the analytic method involving the Laplace transform. On the one hand, the analytic part of the method allows an in-depth understanding of the  physics behind the resonant absorption and phase mixing processes. On the other hand, since the numerical part of the method essentially consists in evaluating analytic expressions, it is much faster than full numerical simulations and is free of the inherent artificial dissipation and resolution limitation  of the numerical codes.  The theory used in this paper is inspired and follows closely that developed by \citet{cally1991}. In essence, we perform  an extension to cylindrical geometry of the planar case of \citet{cally1991}.  The method has also been used in a number of other previous papers \citep[see, e.g.,][]{cally1992,cally1994,mann1995,cally1997,tirry1997} but, to the best of our knowledge, this method has never been used before to study the temporal evolution of  MHD kink waves in  nonuniform cylindrical flux tubes.  

This paper is organized as follows. Section~\ref{sec:model} contains the equilibrium model and the basic equations. The mathematical method is explained in Section~\ref{sec:math}. The process of resonant absorption and  the generation of small spatial scales due to phase mixing  are investigated in Section~\ref{sec:res}. Later, Section~\ref{sec:quasi} is devoted to the damping of the global kink motion and a comparison to the usual global quasi-mode approach is performed. Finally, a discussion on the results and their implications are given in Section~\ref{sec:dis}.

\newpage

\section{MODEL AND BASIC EQUATIONS}
\label{sec:model}

\subsection{Equilibrium}

\begin{figure}
\centering
\includegraphics[width=.4\columnwidth]{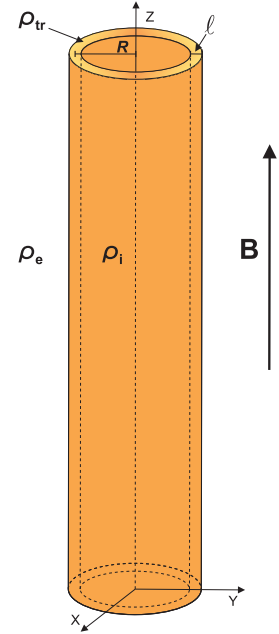}
	\caption{Sketch of the cylindrical flux tube model used in this work.}
	\label{fig:model}
\end{figure}

We consider a simple model for a magnetic flux tube in the solar atmosphere. The sketch of the model is displayed in Figure~\ref{fig:model}. The equilibrium configuration  is composed of  a straight magnetic cylinder of radius $R$ embedded in a uniform and infinite plasma. A cylindrical coordinate system is used, with $r$, $\varphi$, and $z$ representing  the radial, azimuthal, and longitudinal coordinates, respectively. The magnetic field is straight and along the axis of the cylinder, namely ${\bf B} = B {\bf 1}_z$, with $B$ constant. The thermal pressure is constant everywhere. The density, $\rho$, is  uniform in the azimuthal and longitudinal directions, but  nonuniform in the radial direction, namely 
\begin{equation}
 \rho(r) = \left\{
\begin{array}{lll}
\rhoi, & \textrm{if} & r \leq r_1, \\
\rhotr(r), & \textrm{if} & r_1 < r < r_2,\\
\rhoe, & \textrm{if} & r \geq r_2,
\end{array}
\right. \label{eq:density}
\end{equation}
with
\begin{equation}
r_1 \equiv R - \frac{l}{2}, \qquad r_2 \equiv R + \frac{l}{2}.
\end{equation}
We use the subscripts `i' and `e' to denote quantities related to the internal and external plasmas, respectively, and the subscript `tr' to denote quantities related to the transitional layer. In Equation~(\ref{eq:density}), $\rhoi$ and $\rhoe$ are the internal and external constant densities. We set $\rhoi > \rhoe$  to represent an overdense tube. In turn, $\rhotr(r)$ denotes a nonuniform density profile that continuously connects the internal plasma to the external plasma across a layer of thickness  $l\in [0,2R]$. The limit $l= 0$ corresponds to a tube with a piecewise constant density ($r_1=r_2=R$), while the limit $l = 2R$ corresponds to a  fully nonuniform tube in the radial direction ($r_1=0$ and $r_2=2R$).

\subsection{Linear incompressible MHD waves}

 We investigate linear  MHD waves superimposed on the equilibrium magnetic flux tube.  We use the ideal MHD equations \citep[see, e.g.,][]{goedbloed2004,priest2014}. We assume small perturbations over the static background plasma and linearize the ideal MHD equations, i.e., we only keep up to linear terms in the perturbations. The resulting governing equations are
 \begin{eqnarray}
 \rho(r) \frac{\partial^2 \xii}{\partial t^2} &=& - \nabla p' + \frac{1}{\mu} \left( \nabla \times {\bf B}' \right) \times {\bf B}, \label{eq:mom}\\
{\bf B}' &=& \nabla \times \left( \xii \times {\bf B} \right), \label{eq:induc} 
\end{eqnarray}
where $\xii$ is the plasma Lagrangian displacement, ${\bf B}'$ is the magnetic field Eulerian perturbation, $p'$ is the thermal or gas pressure Eulerian perturbation, and $\mu$ is the magnetic permittivity. For simplicity, we shall consider in this work that the plasma perturbations are incompressible. Hence, Equations~(\ref{eq:mom}) and (\ref{eq:induc}) are supplemented with the incompressibility condition, namely
\begin{equation}
\nabla \cdot \xii = 0. \label{eq:inco}
\end{equation}
 We use the incompressibility condition and some vector identities to rewrite  Equation~(\ref{eq:induc}) as
\begin{equation}
{\bf B}' = {\bf B} \cdot \nabla \xii.  \label{eq:induc2} 
\end{equation}
Equation~(\ref{eq:induc2}) allows us to elimate the magnetic field perturbation from the equations. We define the total (thermal + magnetic) pressure perturbation as $P' = p' + {\bf B \cdot B'}/\mu$ and, after some manipulations, we can rewrite Equation~(\ref{eq:mom}) in a compact form, namely
\begin{equation}
\mathcal{L}_{\rm A}  \xii = - \nabla P', 	\label{eq:mom2}
\end{equation}
where $\mathcal{L}_{\rm A}$ is the Alfv\'en operator given by
\begin{equation}
\mathcal{L}_{\rm A} \equiv  \rho(r) \frac{\partial^2 }{\partial t^2} -  \frac{B^2}{\mu}  \frac{\partial^2}{\partial z^2}.
\end{equation}
For later use, we explicitly write the three components of   Equation~(\ref{eq:mom2}), namely
\begin{eqnarray}
 \left( \rho(r) \frac{\partial^2 }{\partial t^2} -  \frac{B^2}{\mu}  \frac{\partial^2}{\partial z^2} \right) \xi_r &=& - \frac{\partial P'}{\partial r}, \label{eq:step1} \\
  \left( \rho(r) \frac{\partial^2 }{\partial t^2} -  \frac{B^2}{\mu}  \frac{\partial^2}{\partial z^2} \right) \xi_\varphi &=& - \frac{1}{r}\frac{\partial P'}{\partial \varphi}, \label{eq:step2} \\
    \left( \rho(r) \frac{\partial^2 }{\partial t^2} -  \frac{B^2}{\mu}  \frac{\partial^2}{\partial z^2} \right) \xi_z &=& -\frac{\partial P'}{\partial z}, \label{eq:step3} 
\end{eqnarray}
where $\xi_r$, $\xi_\varphi$, and $\xi_z$ denote the radial, azimuthal, and longitudinal components of the Lagrangian displacement, respectively. 

Since the equilibrium  is uniform in the $\varphi$- and $z$-directions, we can restrict ourselves to study the individual Fourier components of the perturbations along these directions. Different Fourier components do not interact with each other in the linear regime. Hence the perturbations are put proportional to $\exp(i m \varphi + i k_z z)$, where $m$ and $k_z$ are the azimuthal and longitudinal wavenumbers, respectively.  Only integer values of $m$ are possible. Kink waves are characterized by $m= 1$. These are the only waves that can displace the axis of the flux tube and can move it as a whole.  Concerning the $z$-dependence, note that waves can be either standing or propagating, but the two types are equivalent from the mathematical point of view. From here on, we only retain the explicit dependence of the perturbations on the radial coordinate and time.

 We use  $\xi_r$ as our main variable. For convenience, we  redefine $\xi_\varphi$ and  $\xi_z$  so that they incorporate the imaginary number $i$, namely $i \xi_\varphi \to \xi_\varphi$ and $i \xi_z \to \xi_z $. We do so to avoid the presence of imaginary numbers in the equations. Physically, the factor $i$ accounts for a phase difference of $\pi/2$ with respect to $\xi_r$.  Using Equation~(\ref{eq:inco}) and combining Equations~(\ref{eq:step2}) and (\ref{eq:step3}), we find that the redefined $\xi_\varphi$ and $\xi_z$ are related to $\xi_r$ as
\begin{eqnarray}
 \xi_\varphi &=& - \frac{m/r}{k_z^2 + m^2 / r^2} \frac{1}{r}\frac{\partial \left( r \xi_r \right)}{\partial r}, \label{eq:xif}  \\
 \xi_z &=& - \frac{k_z}{k_z^2 + m^2 / r^2} \frac{1}{r}\frac{\partial \left( r \xi_r \right)}{\partial r}.  \label{eq:xiz}
\end{eqnarray}
In turn, the equation that relates  $P'$ with $\xi_r$ is
\begin{equation}
P' = - \frac{1}{k_z^2 + m^2 / r^2} \left( \rho(r)  \frac{\partial^2}{\partial t^2} +  \frac{B^2}{\mu} k_z^2 \right) \frac{1}{r} \frac{\partial \left( r \xi_r \right)}{\partial r}. \label{eq:prestot}
\end{equation}
We combine Equations~(\ref{eq:step1}) and (\ref{eq:prestot}) to eliminate $P'$ and obtain an equation involving $\xi_r$ only, namely
\begin{equation}
\mathcal{L}_{\rm A}\mathcal{L}_{\rm S}  \xi_r  + \left( k_z^2  + \frac{m^2}{r^2} \right) \frac{\der \rho(r)}{\der r} \frac{\partial^2}{\partial t^2} \frac{1}{r} \frac{\partial \left( r \xi_r \right)}{\partial r} = 0, \label{eq:basicpinit}
\end{equation} 
where the Alfv\'en wave operator, $\mathcal{L}_{\rm A}$, and the surface wave operator, $\mathcal{L}_{\rm S}$,  are   
\begin{eqnarray}
\mathcal{L}_{\rm A} &\equiv & \rho(r)  \frac{\partial^2}{\partial t^2} +  \frac{B^2}{\mu} k_z^2,\\
\mathcal{L}_{\rm S} &\equiv & \left( k_z^2  + \frac{m^2}{r^2} \right) \frac{\partial^2}{\partial r^2} + \left( k_z^2  + \frac{3m^2}{r^2} \right) \frac{1}{r} \frac{\partial}{\partial r} \nonumber \\
&&   - \left[ \left( k_z^2  + \frac{m^2}{r^2} \right)^2 + \left( k_z^2  - \frac{m^2}{r^2} \right) \frac{1}{r^2} \right].
\end{eqnarray}
Equation~(\ref{eq:basicpinit}) governs incompressible MHD waves on the nonuniform flux tube and is the main equation of this investigation.

 When the density is uniform, $\partial \rho(r) / \partial r = 0$ and the second term on the left-hand side of Equation~(\ref{eq:basicpinit}) vanishes. The resulting equation is $\mathcal{L}_{\rm A}\mathcal{L}_{\rm S}  \xi_r =0 $. This equation has two types of  solutions, namely classic Alfv\'en waves and surface MHD waves, which are decoupled when the density is uniform and so they can be studied separately. Readers are referred to \citet{goossens2012} where the properties of both solutions were extensively discussed. On the one hand, the solutions associated with the Alfv\'en wave operator, i.e., the solutions to $\mathcal{L}_{\rm A} \xi_r =0$ correspond to classic Alfv\'en waves. Classic Alfv\'en modes have no motions along the magnetic field direction and have $P' = 0$.  On the other hand, the solutions associated with the surface wave operator, i.e., the solutions to $\mathcal{L}_{\rm S} \xi_r =0$ correspond to surface MHD waves that have $P' \neq 0$. For those modes, $P'$ is maximum at the boundary of the tube and decays far away from the tube.

Conversely, when the density is nonuniform $\partial \rho(r) / \partial r \neq 0$ and the full Equation~(\ref{eq:basicpinit}) must be considered. In such a case, classic Alfv\'en waves and surface MHD waves are no longer separate solutions. We cannot clearly distinguish between classic Alfv\'en waves and surface  MHD waves \citep[see an extensive discussion on this matter in][]{goossens2012}.  The two types of solutions get coupled due to plasma inhomogeneity, whose effect is present in the second term on the left-hand side of Equation~(\ref{eq:basicpinit}). Therefore, the solutions to the full Equation~(\ref{eq:basicpinit}) have mixed properties in the region with nonuniform density, i.e., in the nonuniform boundary layer of the flux tube. In this work we shall see that this fact has a strong impact on the temporal evolution of the plasma motions \citep[see also][]{goossens2014}.

As stated before, we assume that the MHD waves are incompressible. The condition of incompressibility is used  because it greatly simplifies  the mathematical analysis and is  adequate for the type of transverse MHD waves  studied in this paper. It has been shown that in the long-wavelength limit the fundamental radial mode of the MHD kink wave is almost incompressible to a high degree of accuracy \citep[see][]{goossens2009,goossens2012} . Compressibility is of $\mathcal{O}\left(R/\lambda\right)^2$, where $\lambda = 2\pi/ k_z$ is the longitudinal wavelength. The  behavior of the fundamental radial mode is essentially the same in both compressible and incompressible cases  when $R/\lambda \ll 1$ or, equivalently, $k_z R \ll 1$.  The condition of incompressibility also imposes several limitations to our study that must be fairly acknowledged. Due to the condition of incompressibility there are no body MHD waves in the system \citep[see][for the distinction between surface and body waves]{edwin1983}. Body modes produce compression and rarefaction of the plasma in the flux tube and are not included in the analysis. However, this is not a very strong restriction for our work. \citet{goossens2012} showed that the fundamental radial mode of the MHD kink wave has the typical properties of a surface wave in the long-wavelength limit. Also, radial harmonics are absent in the incompressible limit \citep{goossens2012}, and  leaky waves in the external plasma are not possible either \citep[see][]{cally1986}. Both radial harmonics and leaky waves are important in the transitory phase that follows the excitation of standing  oscillations \citep[see][]{terradas2006} and are also involved in the propagation of short-wavelength waves along the tube \citep[see][]{oliver2014}. Hence, to be consistent with the incompressibility assumption, we shall  restrict ourselves to study the fundamental radial mode of long-wavelength  MHD kink waves.

\section{MATHEMATICAL METHOD}
\label{sec:math}

Here we find the solutions to Equation~(\ref{eq:basicpinit}). We stress that we do not assume that there exists a global mode in the flux tube with a certain frequency. Instead, we retain the full temporal dependence of the perturbations. The solutions to Equation~(\ref{eq:basicpinit})  in the regions of the flux tube with uniform and nonuniform density are analyzed separately.

\subsection{Solution in the uniform internal and external regions}
\label{sec:uniform}

As discussed before, classic Alfv\'en waves and surface MHD waves are decoupled when the density in uniform. We are interested in studying the  MHD kink wave, which is a surface mode in the incompressible limit. So, for the sake of simplicity we remove the classic Alfv\'en waves from the scene in the regions with uniform density.  Then, Equation~(\ref{eq:basicpinit}) can be cast as
\begin{eqnarray}
&& \left( k_z^2  + \frac{m^2}{r^2} \right) \frac{\partial^2 \xi_r}{\partial r^2} + \left( k_z^2  + \frac{3m^2}{r^2} \right) \frac{1}{r} \frac{\partial \xi_r}{\partial r} \nonumber \\
&&  - \left[ \left( k_z^2  + \frac{m^2}{r^2} \right)^2 + \left( k_z^2  - \frac{m^2}{r^2} \right) \frac{1}{r^2} \right] \xi_r = 0. \label{eq:modbessel}
\end{eqnarray}
Equation~(\ref{eq:modbessel}) is a differential equation related to the modified Bessel equation \citep[see][]{abramowitz1972}. The solution to Equation~(\ref{eq:modbessel})  is a linear combination of functions $I_m'(k_z r)$ and $K_m'(k_z r)$, where $I_m$ and $K_m$ are the usual modified Bessel functions and the prime $'$ denotes a derivative of the modified Bessel function with respect to its argument.

In the internal plasma, i.e., for $r\leq r_1$, we require $\xi_r$ to be regular at $r=0$. Hence, the solution to Equation~(\ref{eq:modbessel})  in the internal  plasma is
\begin{equation}
\xi_r(r,t) = A_{\rm i}(t) I'_{m}\left( k_z r  \right), \qquad \textrm{if} \qquad r \leq r_1, 
\end{equation}
where $A_{\rm i}(t)$ is a time-dependent amplitude. In turn, in the external plasma, i.e., for $r \geq r_2$, we require $\xi_r$ to vanish when $r\to\infty$, so that solution in the external plasma is
\begin{equation}
\xi_r(r,t) = A_{\rm e}(t) K'_{m}\left( k_z r  \right), \qquad \textrm{if} \qquad r \geq r_2,
\end{equation}
where again $A_{\rm e}(t)$ is a time-dependent amplitude.

We recall that we restrict our analysis to long-wavelength kink waves, i.e., we assume $k_z R \ll 1$ and set $m = 1$. To simplify matters, we perform asymptotic expansions of $I'_{m}$ and $K'_{m}$ for $k_z R \ll 1$, i.e., for small arguments,  and keep the first term in the expansions only.  Therefore, for long-wavelength kink waves the form of $\xi_r$ in the internal and external plasmas simplifies to
\begin{eqnarray}
\xi_r(r,t) & \approx & A_{\rm i}(t), \qquad \textrm{if} \qquad r \leq r_1, \\
\xi_r(r,t) &\approx & A_{\rm e}(t)\, r^{-2}, \qquad \textrm{if} \qquad r \geq r_2,
\end{eqnarray}
where, with no loss of generality, all constants have been absorbed into $A_{\rm i}(t)$ and $A_{\rm e}(t)$.

\subsection{Solution in the nonuniform boundary}

In the nonuniform layer, i.e., for $r_1 < r < r_2$, there is a continuum of Alfv\'en modes that couple to the surface MHD kink wave. We cannot remove Alfv\'en waves from the analysis, and this fact has important physical consequences. We must necessarily consider the full  Equation~(\ref{eq:basicpinit}).  

Following \citet{cally1991}, we preform a  modal expansion of the displacement in the nonuniform boundary of the flux tube. We write $\xi_r(r,t)$ as a generalized Fourier series, namely
\begin{equation}
\xi_r(r,t) = \sum_{n=1}^\infty a_n(t) \psi_n(r), \label{eq:seriessturm}
\end{equation}
where $a_n(t)$ is the time-dependent amplitude of the $n$-th base function $\psi_n(r)$. As appropriate in a cylindrical coordinate system, we choose functions $\psi_n(r)$ to be eigenfunctions of the regular Sturm-Liouville system defined by the Bessel differential equation
\begin{equation}
\frac{\der^2 \psi}{\der r^2} + \frac{1}{r}  \frac{\der \psi}{\der r} + \left( \lambda^2 - \frac{1}{r^2} \right) \psi = 0,
\end{equation}
together with the boundary conditions
\begin{eqnarray}
 \frac{\der \psi}{\der r} &=& 0, \quad \textrm{at} \quad r = r_1,  \\
 \frac{2}{r} \psi + \frac{\der \psi}{\der r} &=& 0, \quad \textrm{at} \quad r = r_2.
\end{eqnarray}
These boundary conditions are chosen for the eigenfunctions to match the spatial behavior of $\xi_r$  at $r=r_1$ and $r=r_2$ (see Section~\ref{sec:uniform}). The eigenfunctions, $\psi_n(r)$, have the following general form,
\begin{equation}
\psi_n(r) = Q_n \left[ J_1 (\lambda_n r) Y'_1 (\lambda_n r_1) - J'_1 (\lambda_n r_1) Y_1 (\lambda_n r)\right],
\end{equation}
where $\lambda_n$ is the  $n$-th eigenvalue, $J_1$ and $Y_1$ are the usual Bessel functions of order 1, and $Q_n$ is a constant determined by imposing the orthogonality condition on the eigenfunctions, namely
\begin{equation}
\frac{1}{l} \int_{r_1}^{r_2} \psi_n(r) \psi_{n'}(r) r \der r = \delta_{n n'}. 
\end{equation}
In turn, the eigenvalues, $\lambda_n$, are the solutions of the transcendental equation, 
\begin{equation}
\frac{\lambda r_2}{2} = - \frac{J_1 (\lambda r_2) Y'_1 (\lambda r_1) - J'_1 (\lambda r_1) Y_1 (\lambda r_2)}{J'_1 (\lambda r_2) Y'_1 (\lambda r_1) - J'_1 (\lambda r_1) Y'_1 (\lambda r_2)}. 
\end{equation}
The constants $Q_n$ and the eigenvalues $\lambda_n$ can be numerically obtained.

\begin{figure}
\centering
\includegraphics[width=.95\columnwidth]{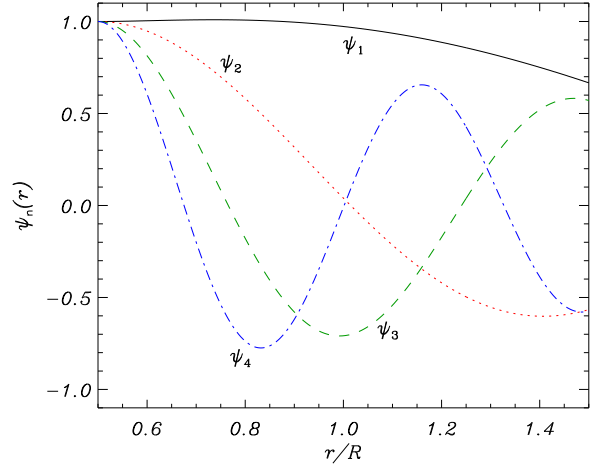}
	\caption{Eigenfunctions of first four generalized Fourier modes in a nonuniform layer with $l/R=1$.  For representation purposes, the eigenfunctions are normalized so that $\psi_n(r_1)=1$.}
	\label{fig:fourier}
\end{figure}

Figure~\ref{fig:fourier} displays the eigenfunctions of the first four generalized Fourier modes. The larger the order of the Fourier mode, the more spatial oscillations of the eigenfunction. In this formalism, large spatial scales are related to  Fourier modes of low order, while small spatial scales are related to  Fourier modes of high order. Therefore,  Fourier modes of low/high order are expected to lose/gain weight as time progresses during the processes of phase mixing and building up of small scales \citep{cally1991,cally1997}. We shall confirm this in Section~\ref{sec:phase}.

Next, we substitute Equation~(\ref{eq:seriessturm}) into Equation~(\ref{eq:basicpinit}) and arrive at a  relation  for the time-dependent amplitudes, $a_n(t)$, namely
\begin{eqnarray}
&& \left[  \rho(r) \mathcal{L}_{\rm S} \psi_{n}(r)  + \frac{\der \rho(r)}{\der r}\left( k_z^2  + \frac{m^2}{r^2} \right) \left( \frac{\der \psi_{n}(r)}{\der r} + \frac{1}{r} \psi_{n}(r) \right)  \right] \frac{\der^2 a_n(t)}{\der t^2} \nonumber \\
&&+  k_z^2 \frac{B^2}{\mu} \mathcal{L}_{\rm S} \psi_{n}(r) a_n(t) = 0, \label{eq:rela}
\end{eqnarray}
for $n=1,2,\dots$. Next, we set the temporal dependence of $a_n$ as $\exp\left( - i \omega t \right)$, multiply Equation~(\ref{eq:rela}) by $\psi_{n'}(r)$, and integrate the resulting equation over the interval $\left[r_1,r_2\right]$ to eliminate the dependence on $r$. Then, a generalized eigenvalue problem can be cast, namely
\begin{equation}
\mathbb{H}\, {\bf a} = \omega^2 \mathbb{M}\,  {\bf a}, \label{eq:eigenproblem}
\end{equation}
where ${\bf a} = [a_1, a_2, \dots]^\mathrm{T}$ is the right eigenvector, $\omega^2$ is the eigenvalue, and $\mathbb{H}$ and $\mathbb{M}$ are square matrices whose elements are
\begin{eqnarray}
H_{nn'} &=& k_z^2 \frac{B^2}{\mu}  \frac{1}{l} \int_{r_1}^{r_2} \psi_{n}(r) \mathcal{L}_{\rm S} \psi_{n'}(r) r \der r , \\
M_{nn'} &=& \frac{1}{l} \int_{r_1}^{r_2} \big[   \rho(r) \mathcal{L}_{\rm S} \psi_{n'}(r)  \nonumber \\
&& \left. + \frac{\der \rho(r)}{\der r}\left( k_z^2  + \frac{m^2}{r^2} \right) \left( \frac{\der \psi_{n'}(r)}{\der r} + \frac{1}{r} \psi_{n'}(r) \right) \right] \psi_{n}(r) r \der r. \nonumber \\
\end{eqnarray}
When the density is uniform, all the eigenvalues are degenerate, namely $\omega^2 = \omegaA^2$, where $\omegaA=k_z \va$ is the Alfv\'en frequency, with $\va = B/\sqrt{\mu \rho}$ the Alfv\'en velocity. When the density is nonuniform the Alfv\'en frequency is spatially-dependent, i.e., a continuum of Alfv\'en frequencies is present. In such a case,  it can be shown  that the eigenvalues are necessarily in the interval $\omegaAi^2 < \omega^2 < \omegaAe^2$, where $\omegaAi = k_z \vai$ and $\omegaAe= k_z \vae$ are the internal and external Alfv\'en frequencies, respectively \citep[see details in][]{cally1991}.

Once the eigenvalue problem  is solved, the time-dependent  amplitudes $a_n$ can be  expressed as a superposition of eigenmodes, namely
\begin{equation}
a_n(t) = \sum_{n'=1}^\infty \beta_{nn'} \left[ c_{n'} \cos\left( \omega_{n'}t \right) + d_{n'} \sin\left( \omega_{n'}t \right) \right],
\end{equation} 
where $\omega_{n'}^2$ is the $n'$-th eigenvalue, $\beta_{nn'}$ is the $n$-th component of the $n'$-th right eigenvector, and ${\bf c} = [c_1, c_2, \dots]^\mathrm{T}$ and ${\bf d} = [d_1, d_2, \dots]^\mathrm{T}$ depend upon the initial conditions as
\begin{eqnarray}
{\bf c} &=&  \mbox{\boldmath{$\mathbb{\beta}$}}^{-1}  {\bf a}(t=0), \label{eq:ces} \\
{\bf d} &=&  \mbox{\boldmath{$\mathbb{\beta}$}}^{-1} \left. \frac{\der {\bf a}}{\der t}\right|_{t=0},
\end{eqnarray}
where \mbox{\boldmath{$\mathbb{\beta}$}} is a matrix whose elements are $\beta_{nn'}$. The components of ${\bf a}$ and their derivatives at $t=0$ are computed from the initial conditions for $\xi_r$ as 
\begin{eqnarray}
a_n(t=0) &=& \int_{r_1}^{r_2} \xi_r(r,t=0) \psi_n(r) r \der r, \\
 \left. \frac{\der a_n}{\der t}\right|_{t=0} &=& \int_{r_1}^{r_2} \left.\frac{\partial \xi_r(r,t)}{\partial t}\right|_{(r,t=0)} \psi_n(r) r \der r.
\end{eqnarray}
Finally, the full expression of $\xi_r(r,t)$ in the nonuniform layer is
\begin{equation}
\xi_r(r,t) = \sum_{n=1}^\infty \sum_{n'=1}^\infty \beta_{nn'} \left[ c_{n'} \cos\left( \omega_{n'}t \right) + d_{n'} \sin\left( \omega_{n'}t \right) \right] \psi_n(r). \label{eq:pmodal}
\end{equation}

\begin{figure*}
\centering
\includegraphics[width=.95\columnwidth]{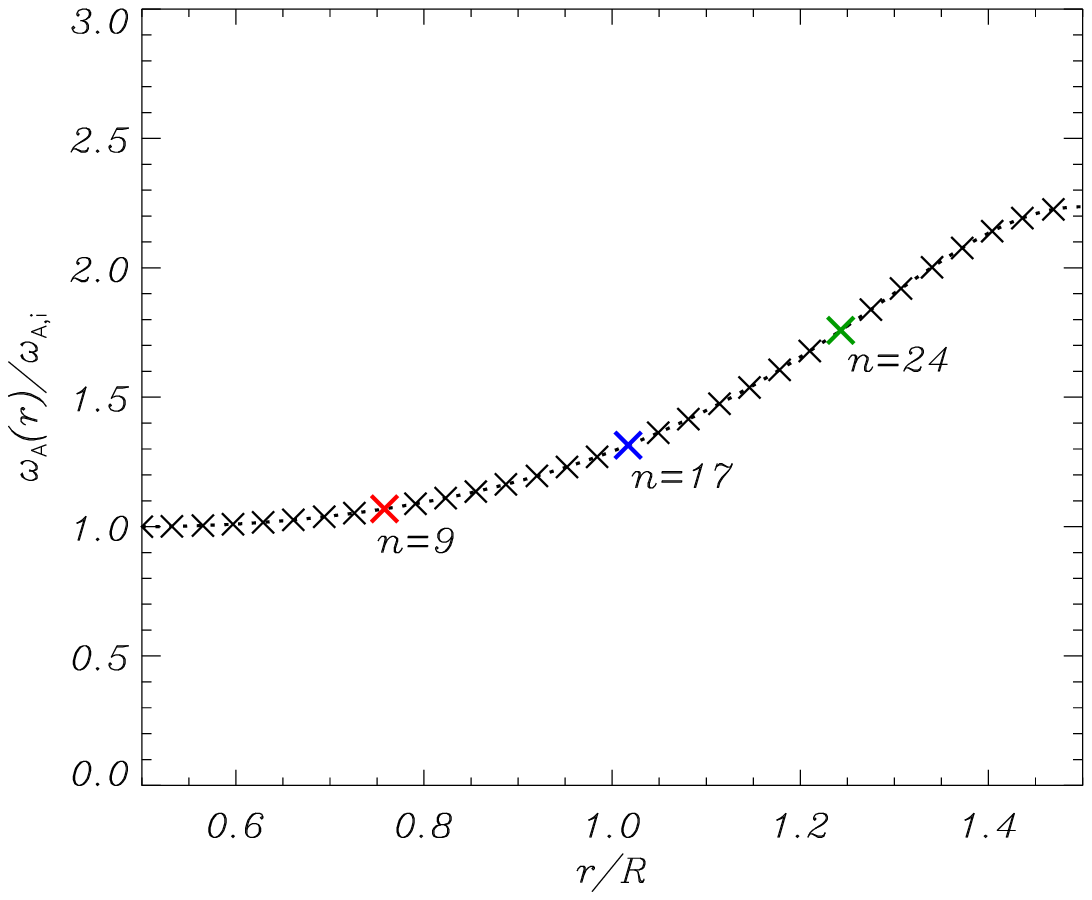}
\includegraphics[width=.95\columnwidth]{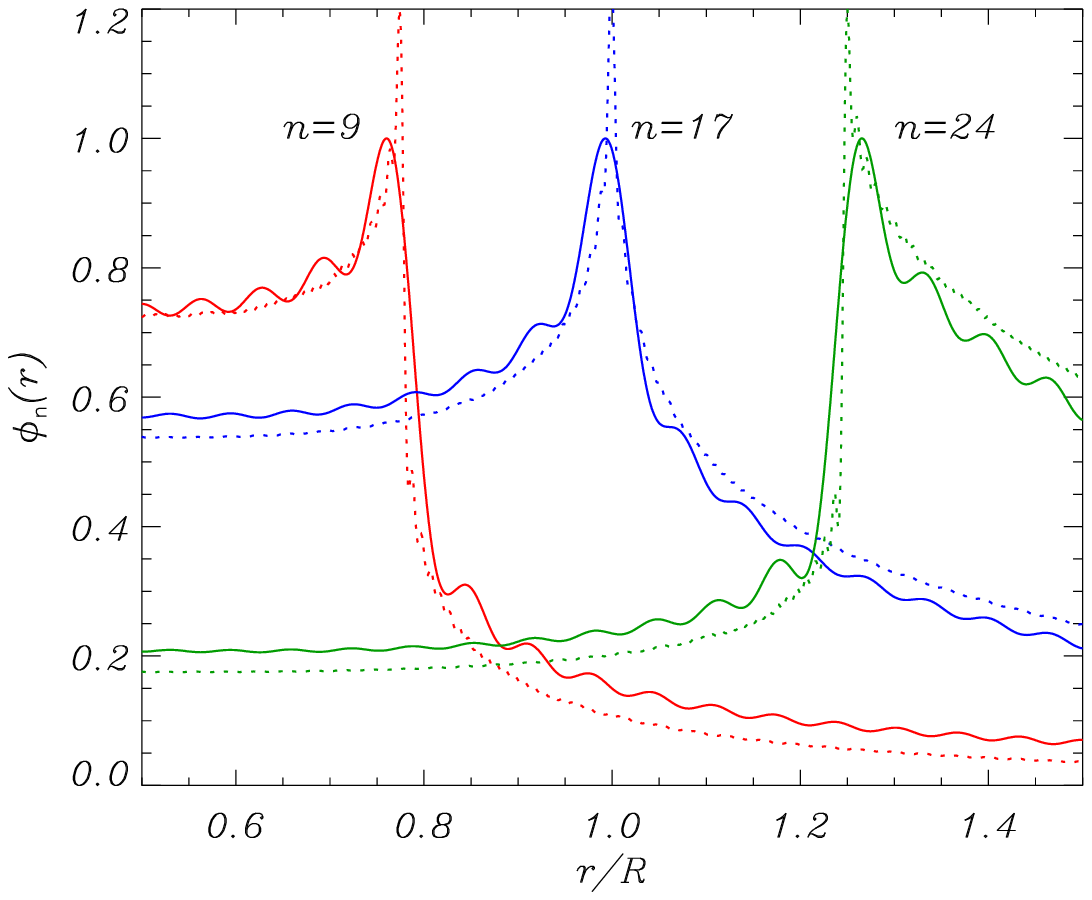}
	\caption{Left: Alfv\'en contiuum of frequencies (dotted line) as function of position in the nonuniform layer. The superimposed crosses correspond to the discrete eigenfrequencies for $N=31$. Right: Eigenfunctions of three selected discrete modes for $N=31$ (solid lines) and for $N=201$ (dotted lines). In the computations we used a sinusoidal transition of density, $\rhoi/\rhoe=5$, $l/R=1$, $m=1$, and $k_zR = \pi/100$.}
	\label{fig:barston}
\end{figure*}

\subsection{Alfv\'en continuum modes}

The generalized eigenvalue problem of Equation~(\ref{eq:eigenproblem}) involve infinite matrices. To solve the  eigenvalue problem numerically, we must truncate the generalized Fourier series so that only the first $N$ Fourier modes are considered. Hence, we must replace $\infty$ by $N$ in the upper limit of all summations. Then, the infinite matrices become finite $N \times N$ matrices and, consequently, the number of eigenvalues is $N$.  Furthermore, in order to shed light on the physics behind  Equation~(\ref{eq:pmodal}), we recast  it in the following alternative form,
\begin{equation}
\xi_r(r,t) = \sum_{n=1}^N \left[ c_{n} \cos\left( \omega_{n}t \right) + d_{n} \sin\left( \omega_{n}t \right) \right] \phi_{n}(r), \label{eq:pmodalalt}
\end{equation}
with
\begin{equation}
\phi_{n}(r) \equiv \sum_{n'=1}^N \beta_{n'n} \psi_{n'}(r).
\end{equation}
As explained by \citet{cally1991},  functions $\phi_{n}(r)$, and their associated  frequencies $\omega_n$, play the role of the Alfv\'en contiuum modes.  The whole set of these $N$ modes represents a discretized version of the Alfv\'en contiuum. Equation~(\ref{eq:pmodalalt}) is  useful because it allows us to physically understand that the total displacement, $\xi_r(r,t)$, is the superposition of the displacements produced by all the modes of the Alfv\'en continuum. These discrete `Alfv\'en contiuum modes' are equivalent to the modes discussed by \citet{barston1964} in the context of electrostatic oscillations in a nonuniform plasma. As shown by \citet{barston1964}, no global  mode is present when the density is nonuniform.

Figure~\ref{fig:barston} (left panel) displays the Alfv\'en continuum of frequencies, $\omegaA(r)$, as function of position in the nonuniform layer for a particular set of parameters given in the caption of the Figure. In this computation we used a sinusoidal variation of density in the nonuniform layer, namely
\begin{equation}
\rho_{\rm tr}(r) = \frac{\rhoi}{2} \left[\left( 1 + \frac{\rhoe}{\rhoi} \right) -  \left( 1 - \frac{\rhoe}{\rhoi} \right) \sin \left(\frac{\pi}{l}(r-R) \right) \right]. \label{eq:sin}
\end{equation}
We overplot in Figure~\ref{fig:barston} (left panel) the discrete eigenfrequecies, $\omega_n$, that are solutions to the truncated eigenvalue problem  using $N=31$. We see that the Alfv\'en contiuum is correctly recovered by the discrete eigenfrequencies. We have selected three of these discrete modes and have plotted their  eigenfunctions in the right panel of Figure~\ref{fig:barston}. This Figure can be compared to Figure~8 of \citet{cally1991} for equivalent eigenfunctions in planar geometry. The eigenfunctions are discontinuous (singular) at the specific position in the nonuniform layer where the discrete  eigenfrequency matches the local Alfv\'en frequency. We notice both logarithmic and step discontinuities in the plotted eigenfunctions, as consistent with the expected behavior of the Alfv\'en continuum modes \citep[see details in][]{goedbloed2004}. The  singularities are not well resolved in the right panel of Figure~\ref{fig:barston} because, on purpose, we considered a relatively small value of $N$. As a consequence, there are too few regular Fourier modes to accurately capture the singular behavior of the eigenfunctions. Also note the presence of  wobbles as a sort of Gibbs phenomenon in the eigenfunctions. These issues can easily be overcome by increasing the value of $N$. The larger the value of $N$, the larger the number of Fourier modes and so the better representation of the  contiuum modes. For instance, we have increased the number of Fourier modes to $N=201$ and have computed again the eigenfunctions of the discrete `Alfv\'en contiuum modes' whose frequencies are the same as those displayed in  Figure~\ref{fig:barston} (right panel). We have overplotted these new eigenfunctions using dotted lines. The singular behavior of the eigenfunctions is now well captured and the  wobbles are much less noticeable than for $N=31$.

Note in Figure~\ref{fig:barston} (right panel) that the `wings' of the eigenfunctions penetrate in the internal and external plasmas. Hence, in the regions with uniform density the added contribution of the `wings' of all the  continuum modes are responsible for the total  displacement. In addition, the fact that every  continuum mode has a different frequency has the consequence that the oscillations in the nonuniform layer will eventually get out of phase as time progresses. In essence, this is the process of phase mixing \citep[see, e.g.,][]{pritchett1978,heyvaerts1983}. A question then arises: what happens to the displacement in the internal and external regions with uniform density as phase mixing occurs in the nonuniform layer? We anticipate that the net result is the apparent damping of the global kink motion. This is explored in detail in Section~\ref{sec:res}.

\section{TEMPORAL EVOLUTION OF THE DISPLACEMENT}
\label{sec:res}

Here, we show the result of the temporal evolution of the displacement components in the flux tube  using the modal expansion method. As initial condition for the radial component of the displacement we set 
\begin{equation}
\xi_r(r,t=0) = \left\{ \begin{array}{lll}
 \xi_0 , & \textrm{if} & r \leq r_1, \\
\xi_0 \frac{\psi_1 (r)}{\psi_1 (r_1)}, & \textrm{if} & r_1 < r < r_2, \\
\xi_0 \frac{\psi_1 (r_2)}{\psi_1 (r_1)}\left( \frac{r_2}{r} \right)^2, & \textrm{if} & r \geq r_2,
\end{array} \right. \label{eq:initial}
\end{equation}
and $\left. \partial \xi_r / \partial t \right|_{(r,t=0)}=0$ for all $r$, where $\xi_0$ is the arbitrary internal amplitude at $t=0$. The azimuthal, $\xi_\varphi$, and longitudinal, $\xi_z$, components of the displacement depend on $\xi_r$ through Equations~(\ref{eq:xif}) and (\ref{eq:xiz}), respectively.  With this initial condition,  only the first generalized Fourier mode, i.e., the mode with the largest spatial scale, contributes to the total displacement at $t=0$. We choose this initial condition on purpose to investigate the process of building up of small scales that are related to Fourier modes of large order. In all computations we use $N=201$ to assure a good accuracy of the modal scheme. Unless otherwise stated, we consider a sinusoidal transition of density (Equation~(\ref{eq:sin})) and use the following set of parameters: $\rhoi/\rhoe = 5$, $m=1$, and $k_zR = \pi/100$.

\begin{figure}
\centering
\includegraphics[width=.95\columnwidth]{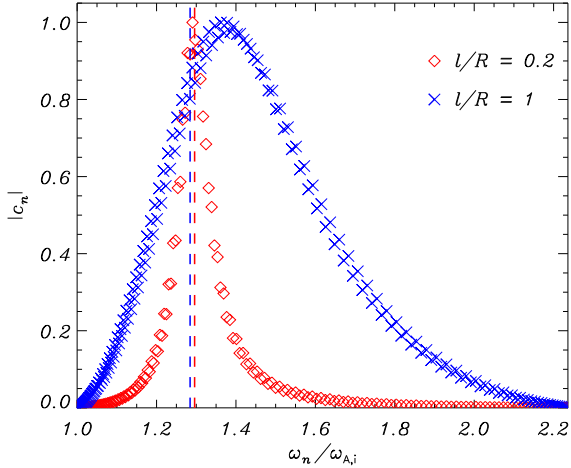}
	\caption{Normalized amplitudes of the discrete `Alfv\'en contiuum modes' as function of their frequencies according to the initial condition given in Equation~(\ref{eq:initial}). The red diamonds correspond to $l/R=0.2$ and the blue crosses are for $l/R=1$. The vertical dashed lines denote the corresponding quasi-mode frequencies.}
	\label{fig:cs}
\end{figure}

First of all, we solve the generalized eigenvalue problem (Equation~(\ref{eq:eigenproblem})) for two  values of the nonuniform layer thickness, namely a thin layer with $l/R = 0.2$  and a thick layer with $l/R=1$. Then, we compute the amplitudes, $c_n$, of the discrete `Alfv\'en contiuum modes' using the initial condition given in Equation~(\ref{eq:initial}). Note that $d_n = 0$ since $\left. \partial \xi_r / \partial t \right|_{(r,t=0)}=0$. The normalized values of $c_n$ are plotted against their corresponding eigenfrequencies in Figure~\ref{fig:cs}. In the case of a thin layer, we find a sharp peak  around a specific frequency, which indicates that the `Alfv\'en contiuum modes' that  mostly contribute to the total displacement are those whose frequencies are near that specific frequency. However, in the case of a thick layer a broader spectrum of modes contributes to the total displacement.  We shall discuss in Section~\ref{sec:quasi} the relation between the position and width of the peak in the frequency spectrum and the so-called quasi-mode frequency and damping rate \citep[see also][]{andries2007}.

Next, the temporal evolution of the displacement components is computed. Figures~\ref{fig:res1} and \ref{fig:res2} display the temporal evolution  for  $l/R = 0.2$ (Figure~\ref{fig:res1}) and  $l/R=1$ (Figure~\ref{fig:res2}). See also the accompanying animations\footnote{Movies can be downloaded from \url{http://www.uib.es/depart/dfs/Solar/movies_SolerTerradas2015.zip}}.   Time is expressed in units of $P_k = 2\pi/\omega_k$, where $P_k$ is the kink mode period of a homogeneous thin tube, with  $\omega_k$  the so-called kink frequency \citep[see, e.g,][]{edwin1983} given by
\begin{equation}
\omega_k = \sqrt{\frac{\rhoi\omegaAi^2 + \rhoe\omegaAe^2}{\rhoi + \rhoe}}. \label{eq:wk}
\end{equation} 
In the following subsections, we discuss in detail the dynamics shown in  Figures~\ref{fig:res1} and \ref{fig:res2}. These results can be compared to those obtained from  full numerical simulations of kink oscillations \citep[e.g.,][]{terradas2006,terradas2008}. Also, although found in a very different context, our results show many similarities to those obtained by \citet{mann1995} in the case of resonant waves in the magnetosphere.

\subsection{Resonant absorption and flux of energy to the nonuniform boundary}
\label{sec:resonant}

\begin{figure*}
\centering
\includegraphics[width=.62\columnwidth]{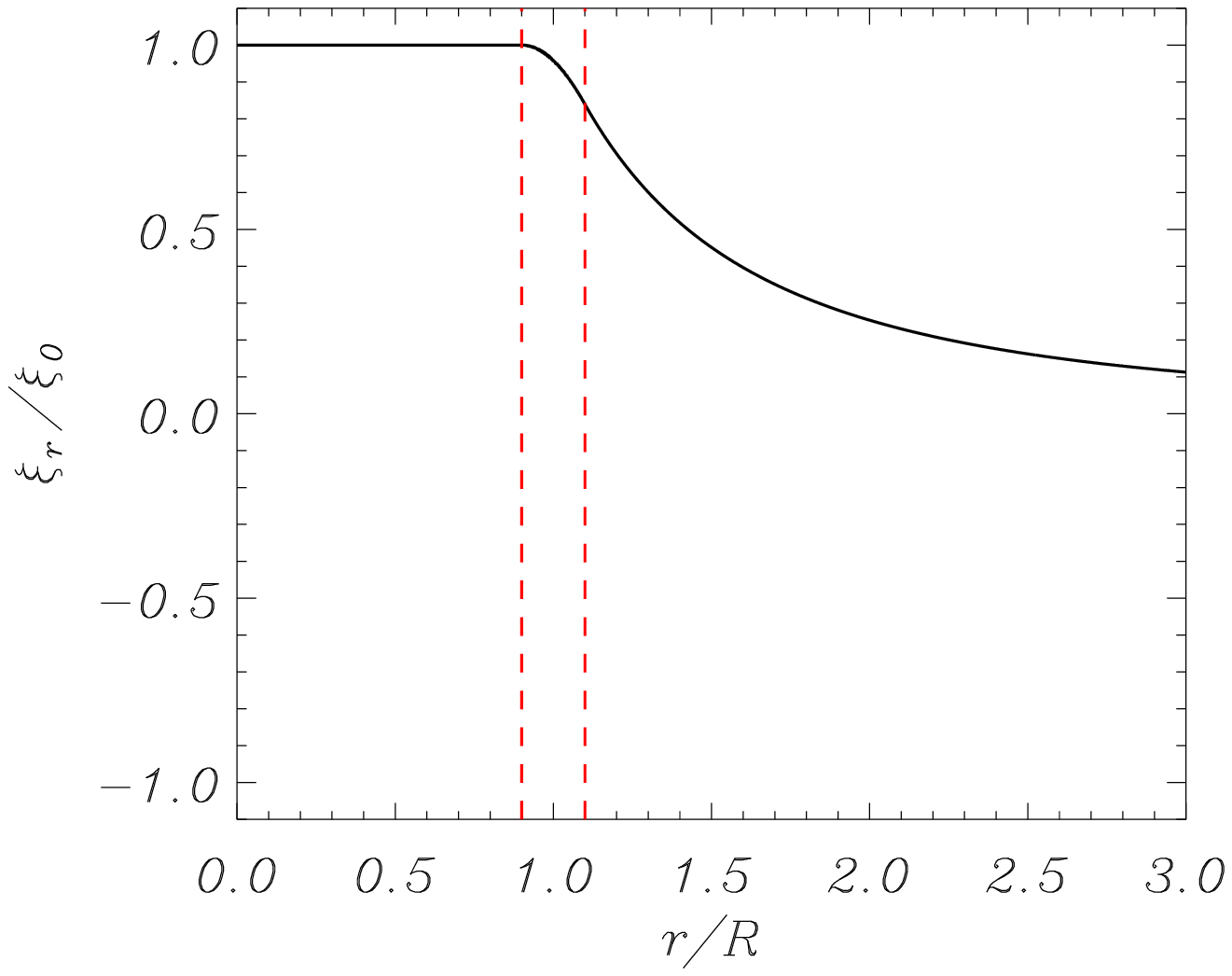}
\includegraphics[width=.62\columnwidth]{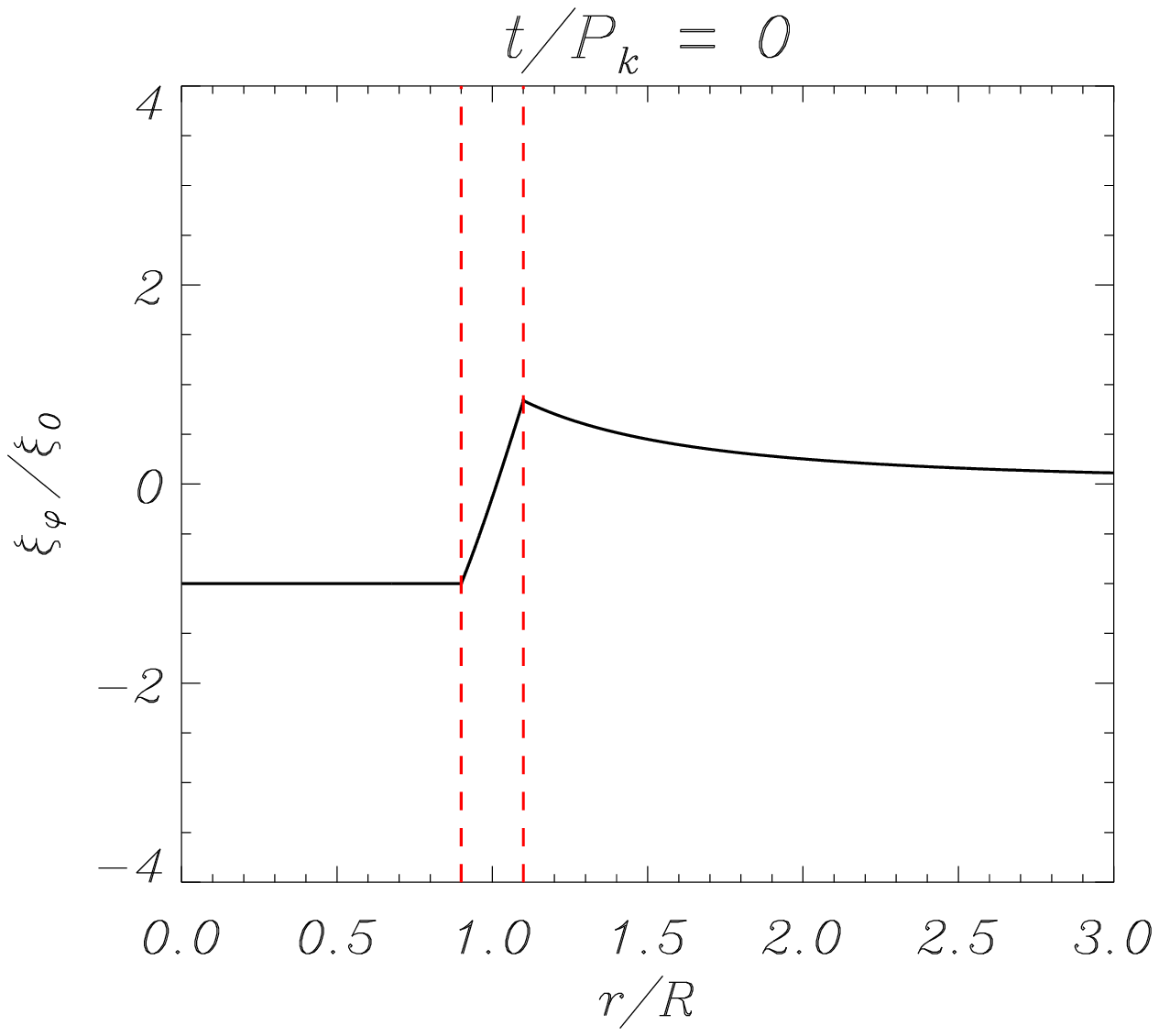}
\includegraphics[width=.62\columnwidth]{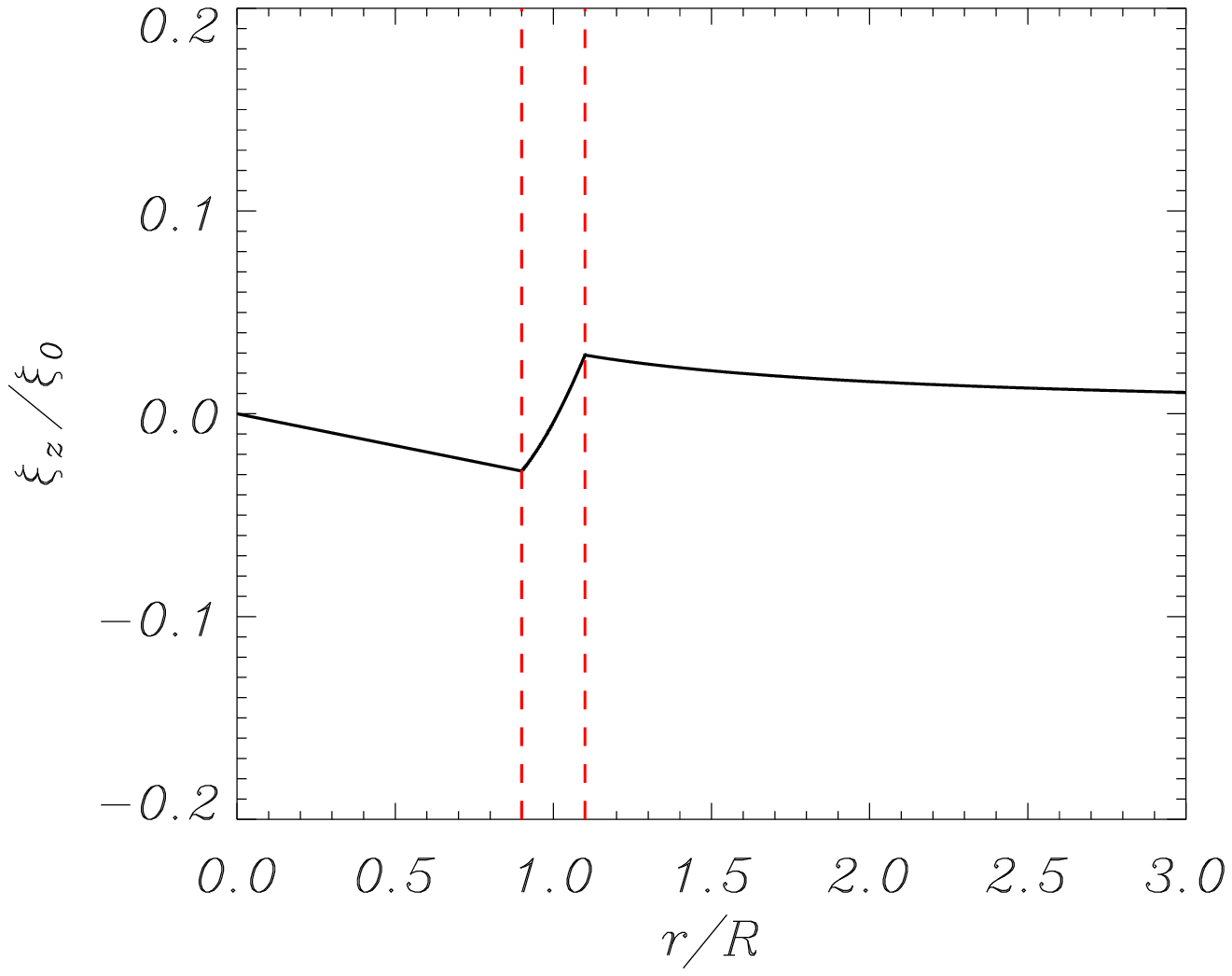}
\includegraphics[width=.62\columnwidth]{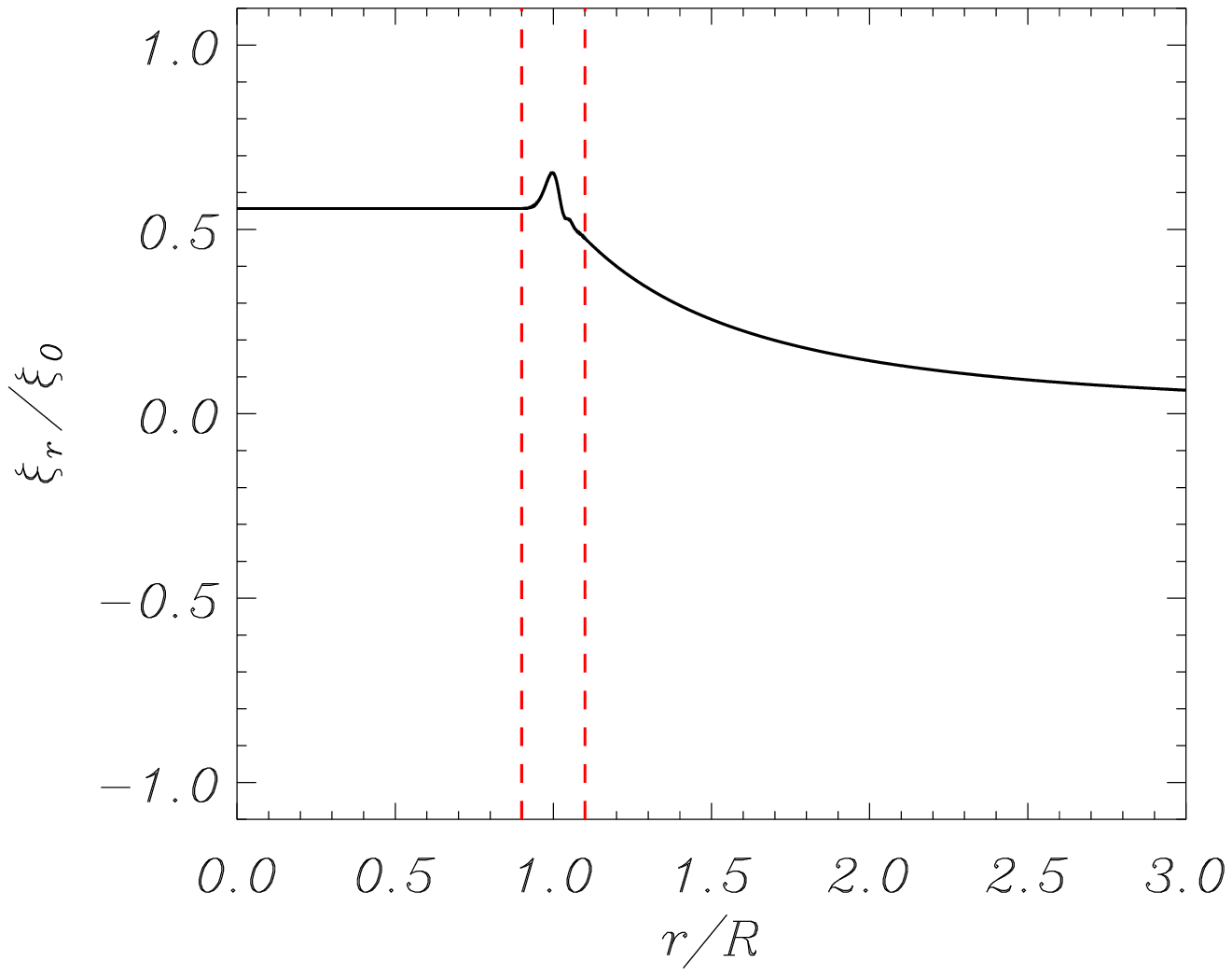}
\includegraphics[width=.62\columnwidth]{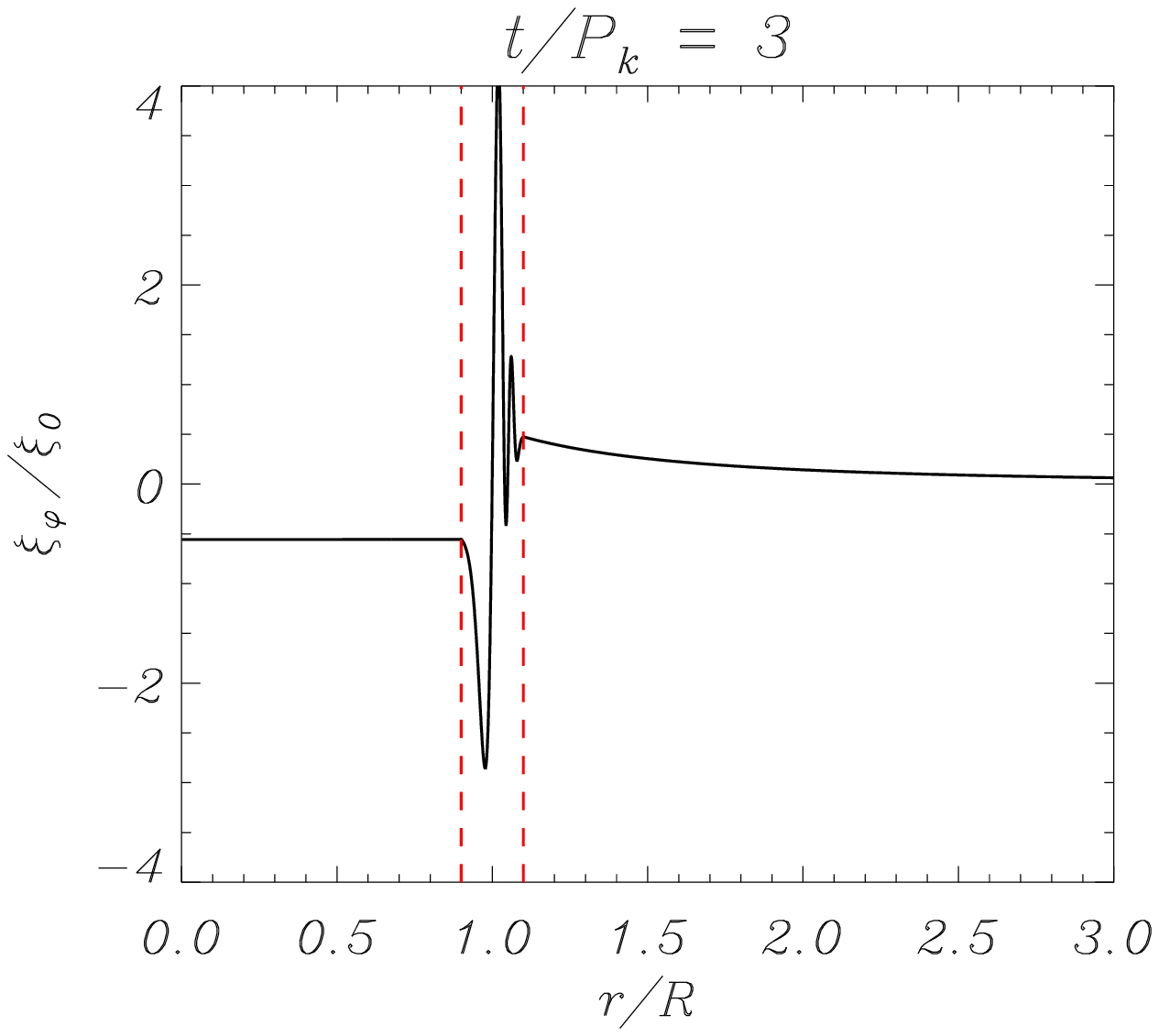}
\includegraphics[width=.62\columnwidth]{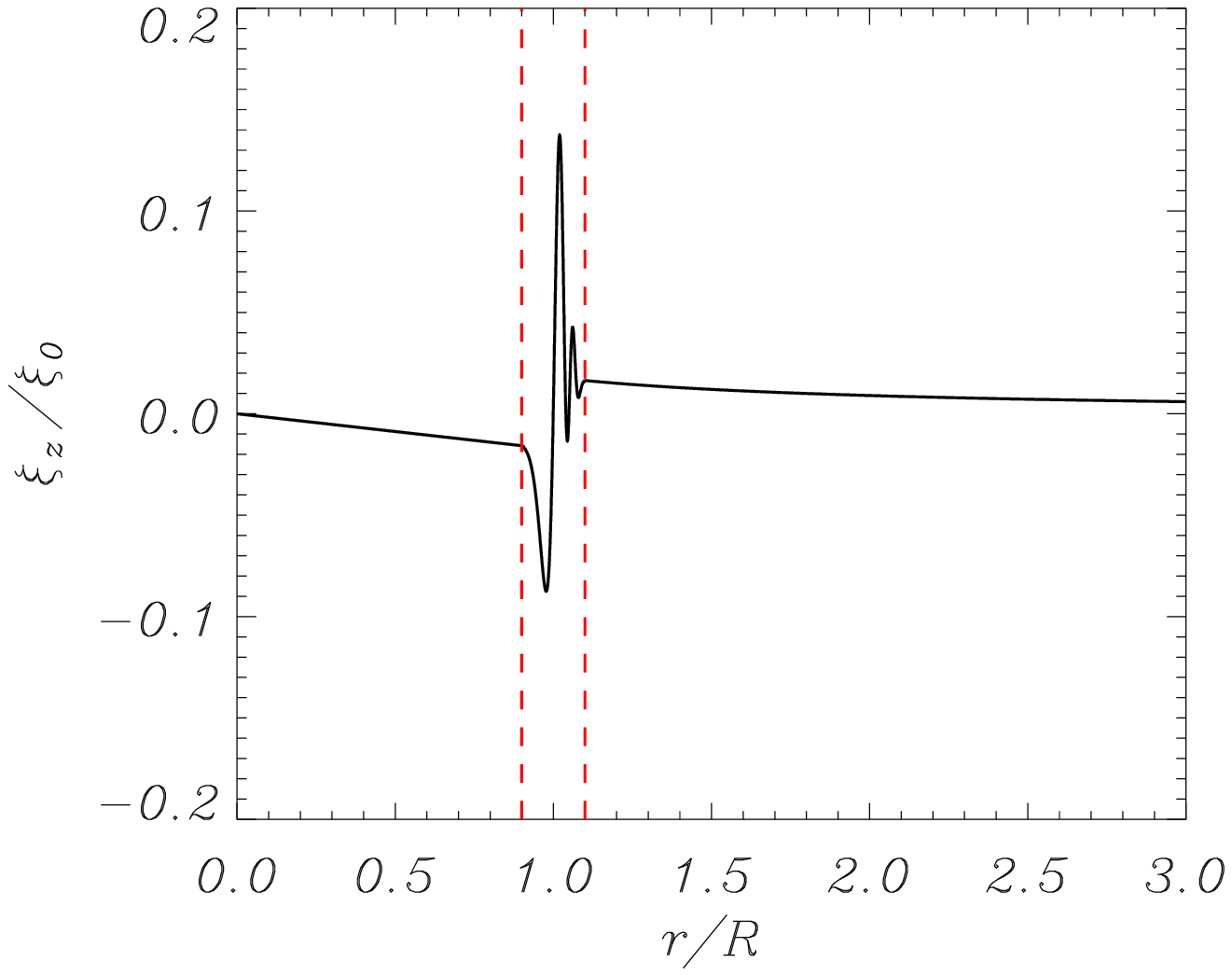}
\includegraphics[width=.62\columnwidth]{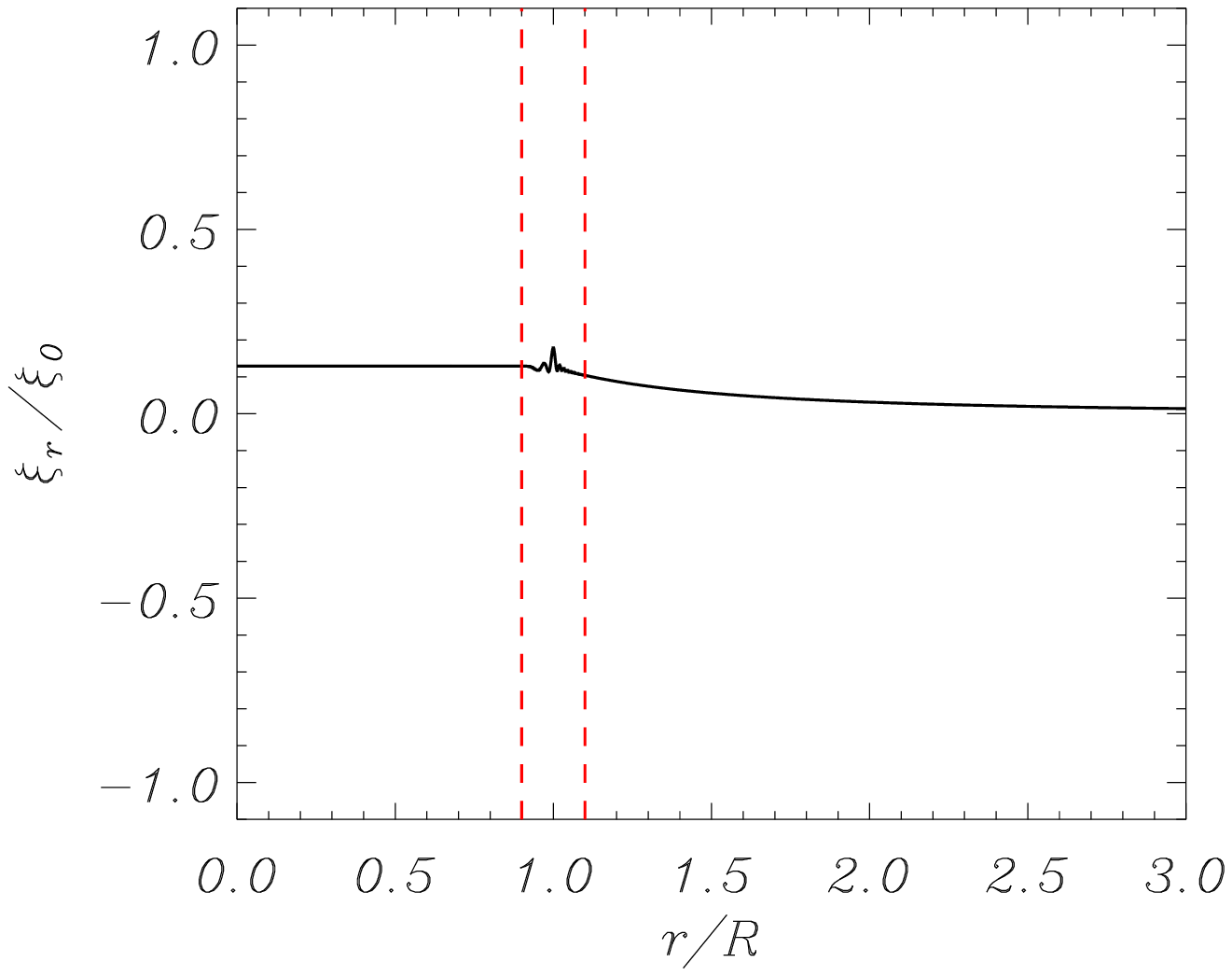}
\includegraphics[width=.62\columnwidth]{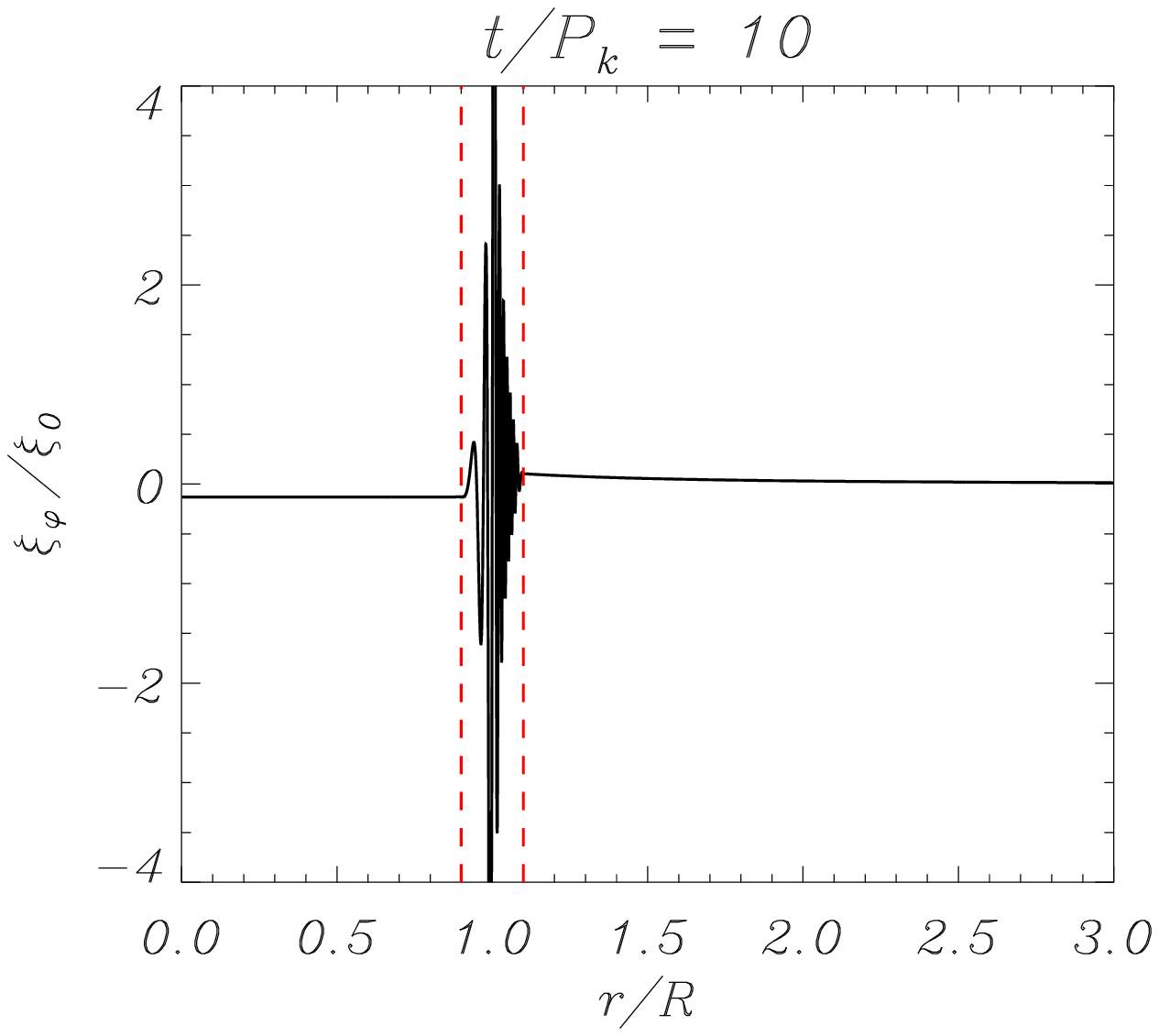}
\includegraphics[width=.62\columnwidth]{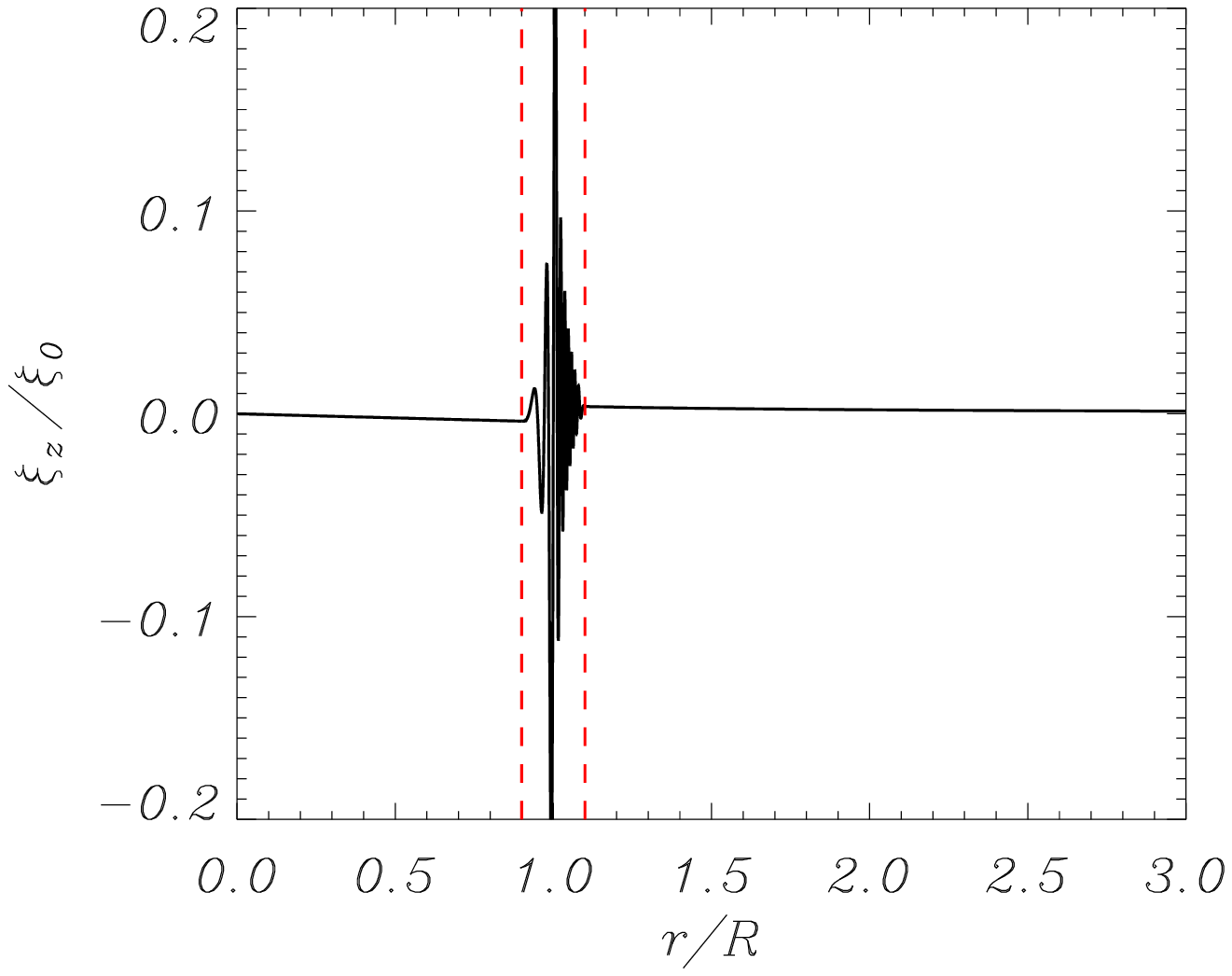}
	\caption{Temporal evolution of $\xi_r$ (left), $\xi_\varphi$ (center), and $\xi_z$ (right) in a nonuniform tube with $l/R=0.2$. Top, mid, and bottom panels are for $t/P_k =$~0, 3, and 10, respectively. The  vertical dashed lines denote the boundaries of the nonuniform layer. Note that the vertical scale is different in each panel. Movie can be downloaded from \url{http://www.uib.es/depart/dfs/Solar/movies_SolerTerradas2015.zip} }
	\label{fig:res1}
\end{figure*}

\begin{figure*}
\centering
\includegraphics[width=.62\columnwidth]{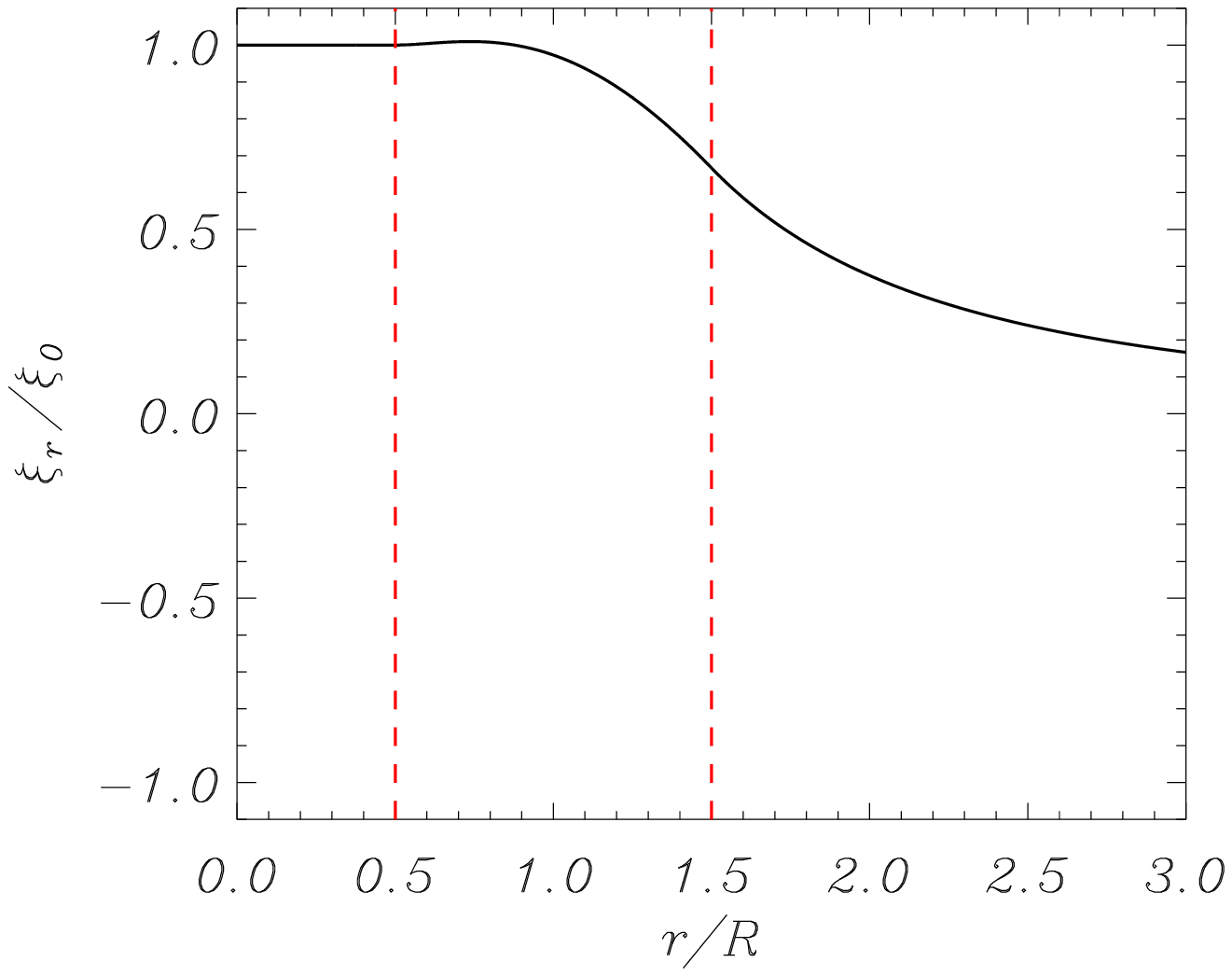}
\includegraphics[width=.62\columnwidth]{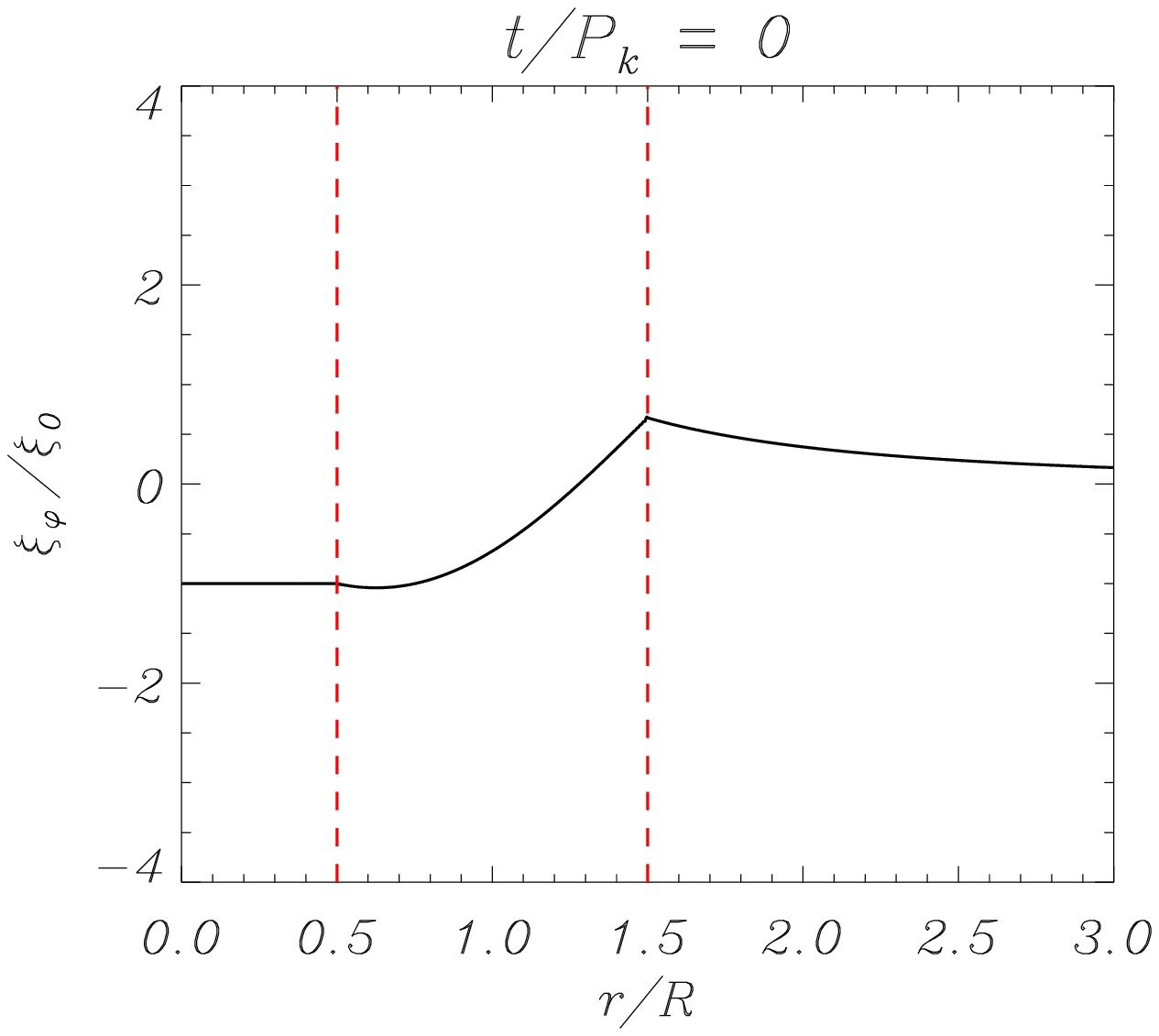}
\includegraphics[width=.62\columnwidth]{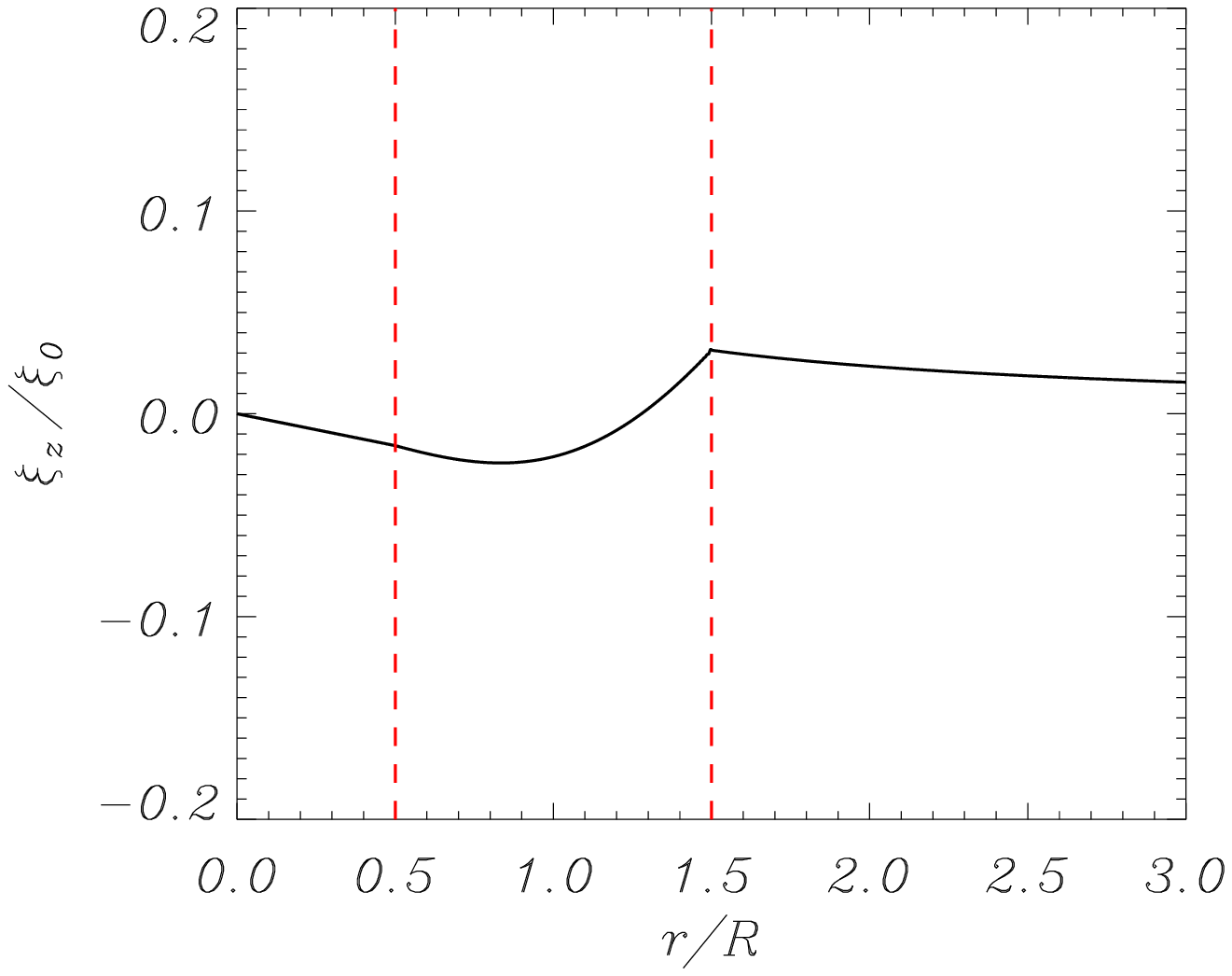}
\includegraphics[width=.62\columnwidth]{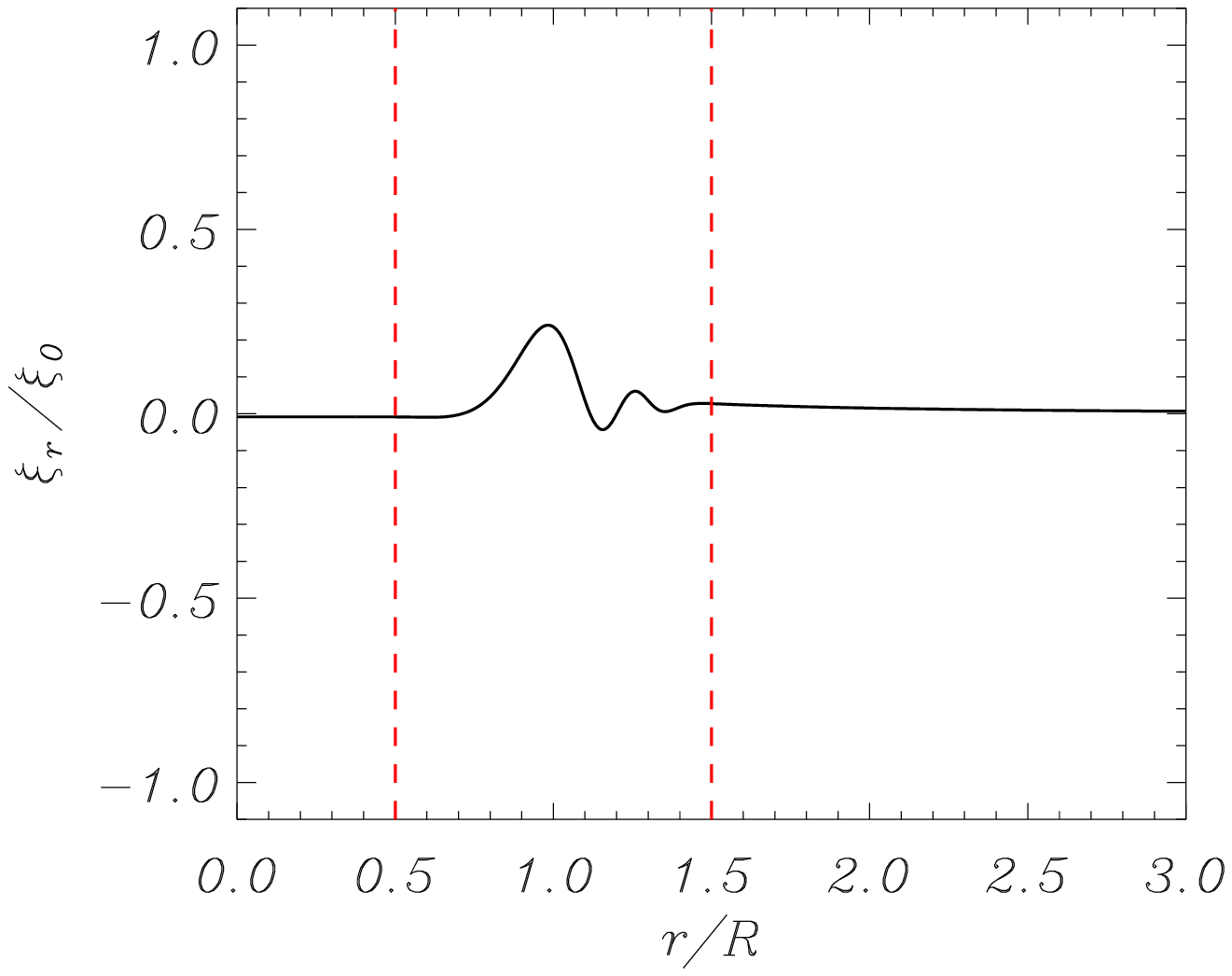}
\includegraphics[width=.62\columnwidth]{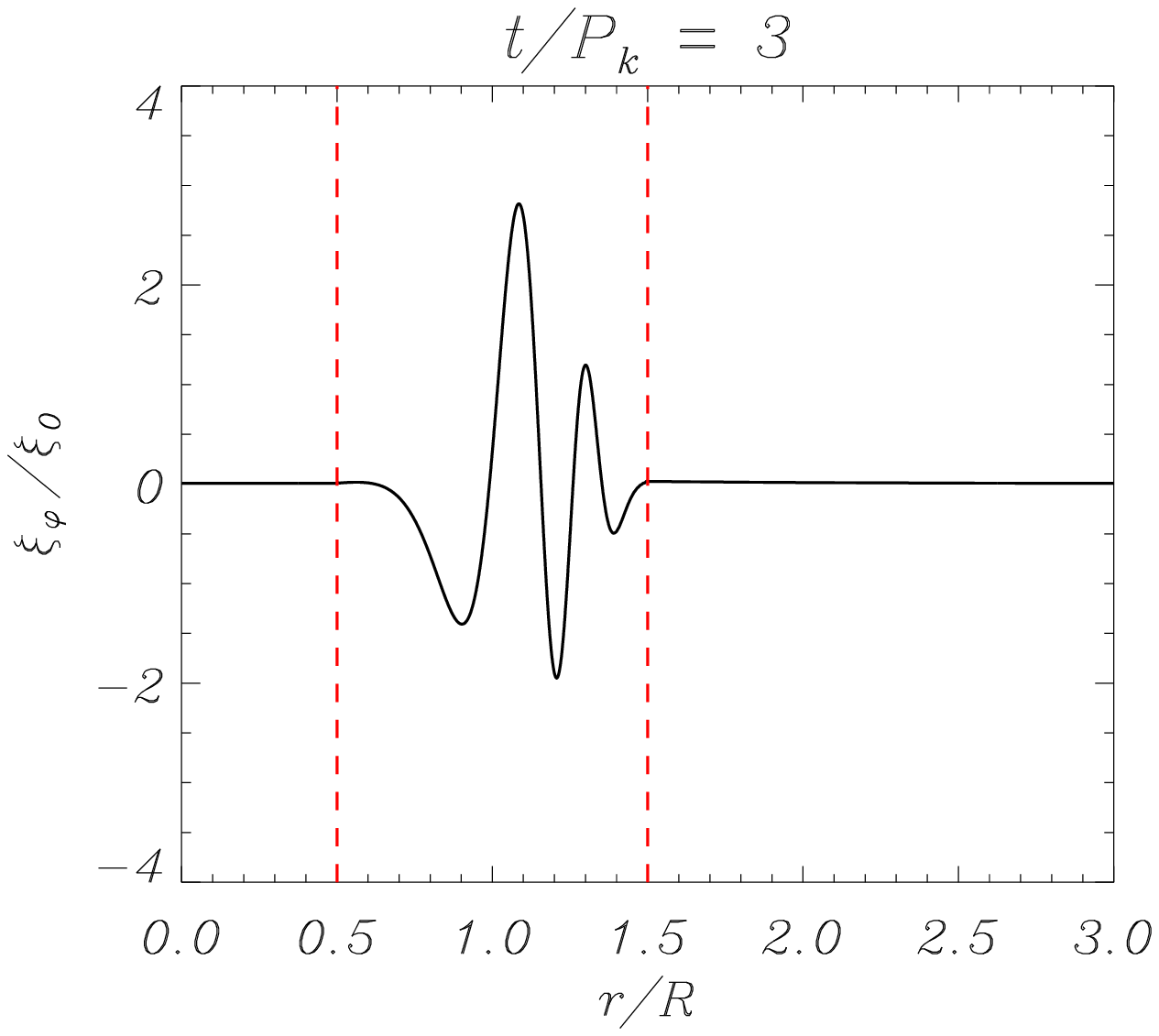}
\includegraphics[width=.62\columnwidth]{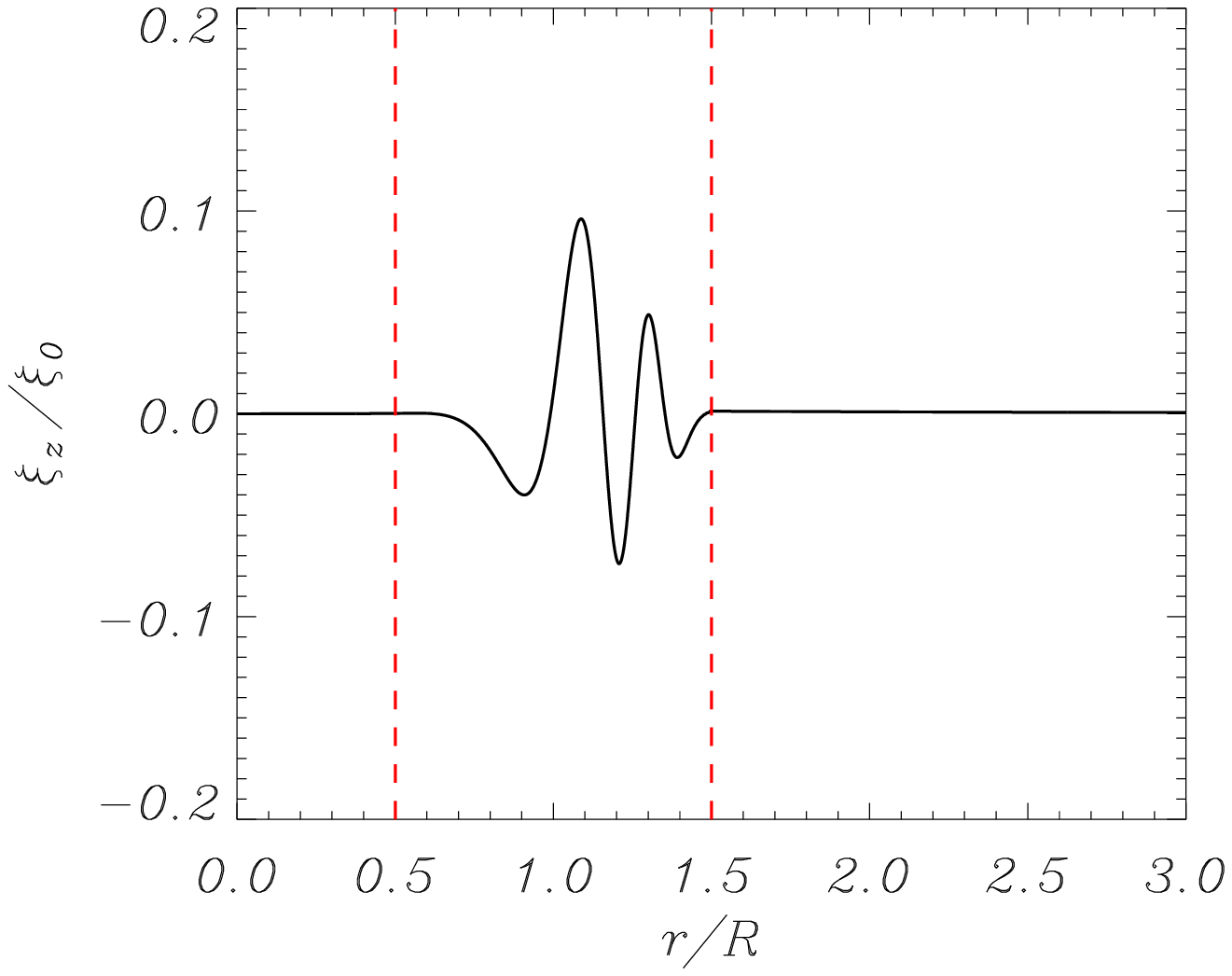}
\includegraphics[width=.62\columnwidth]{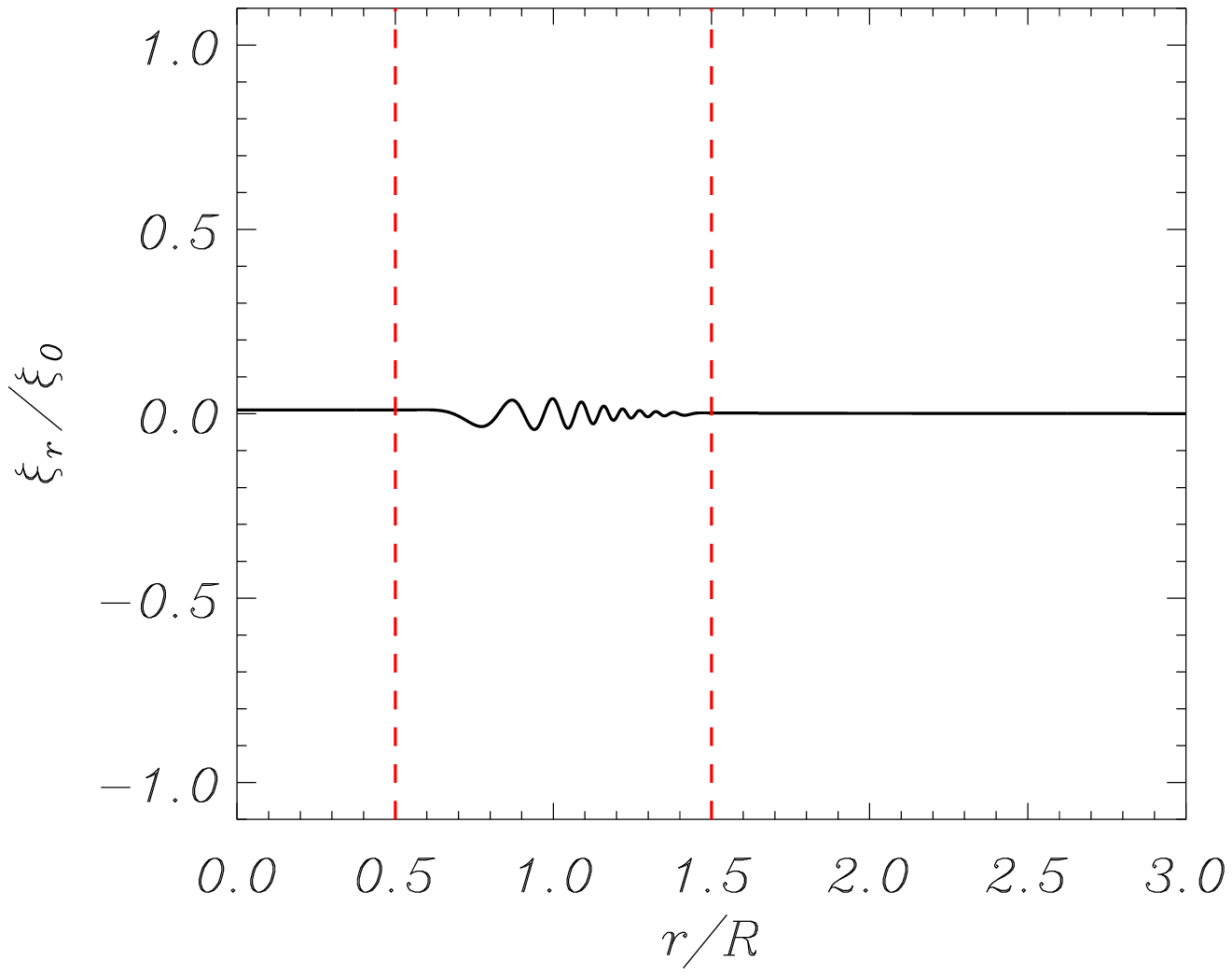}
\includegraphics[width=.62\columnwidth]{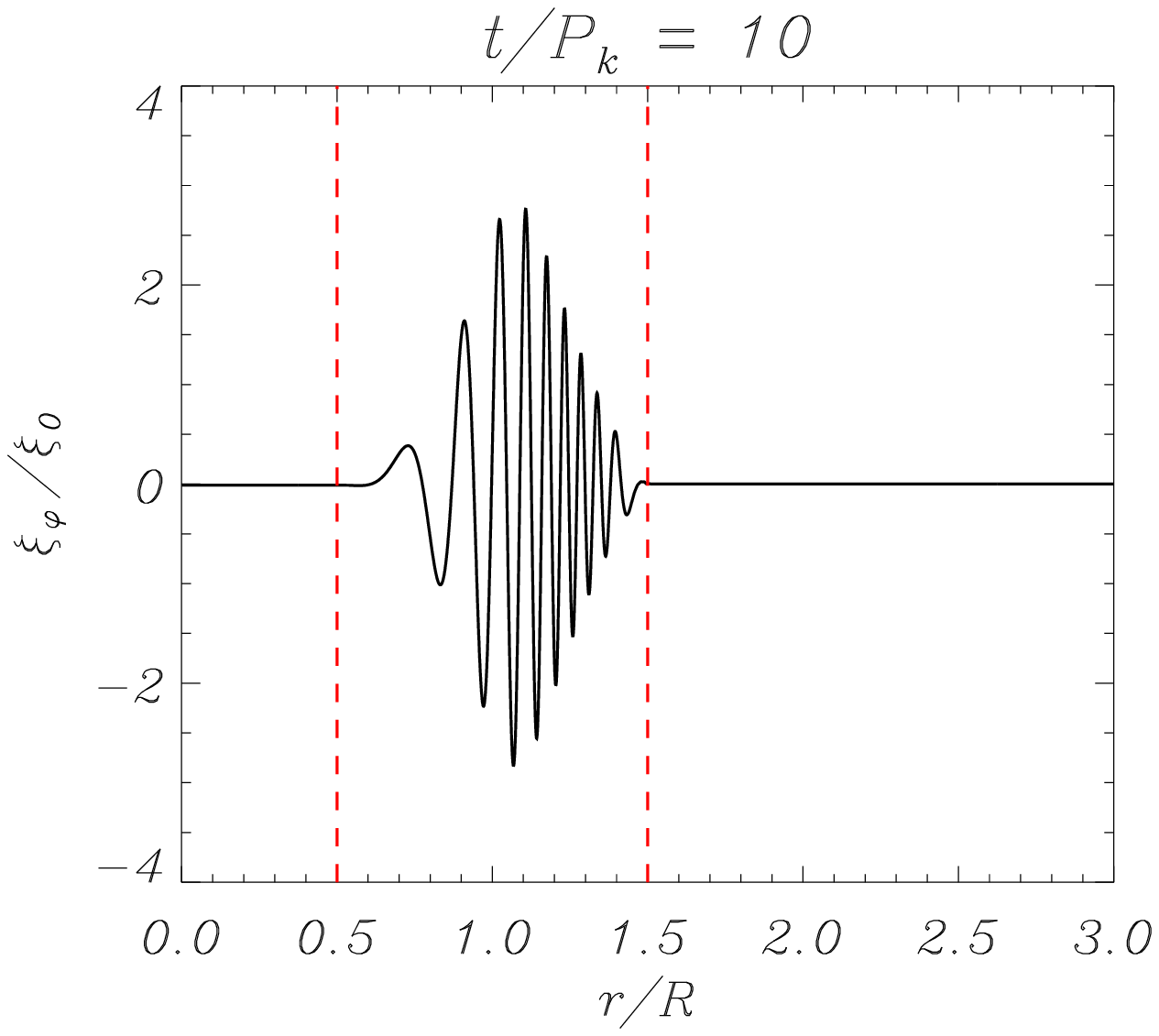}
\includegraphics[width=.62\columnwidth]{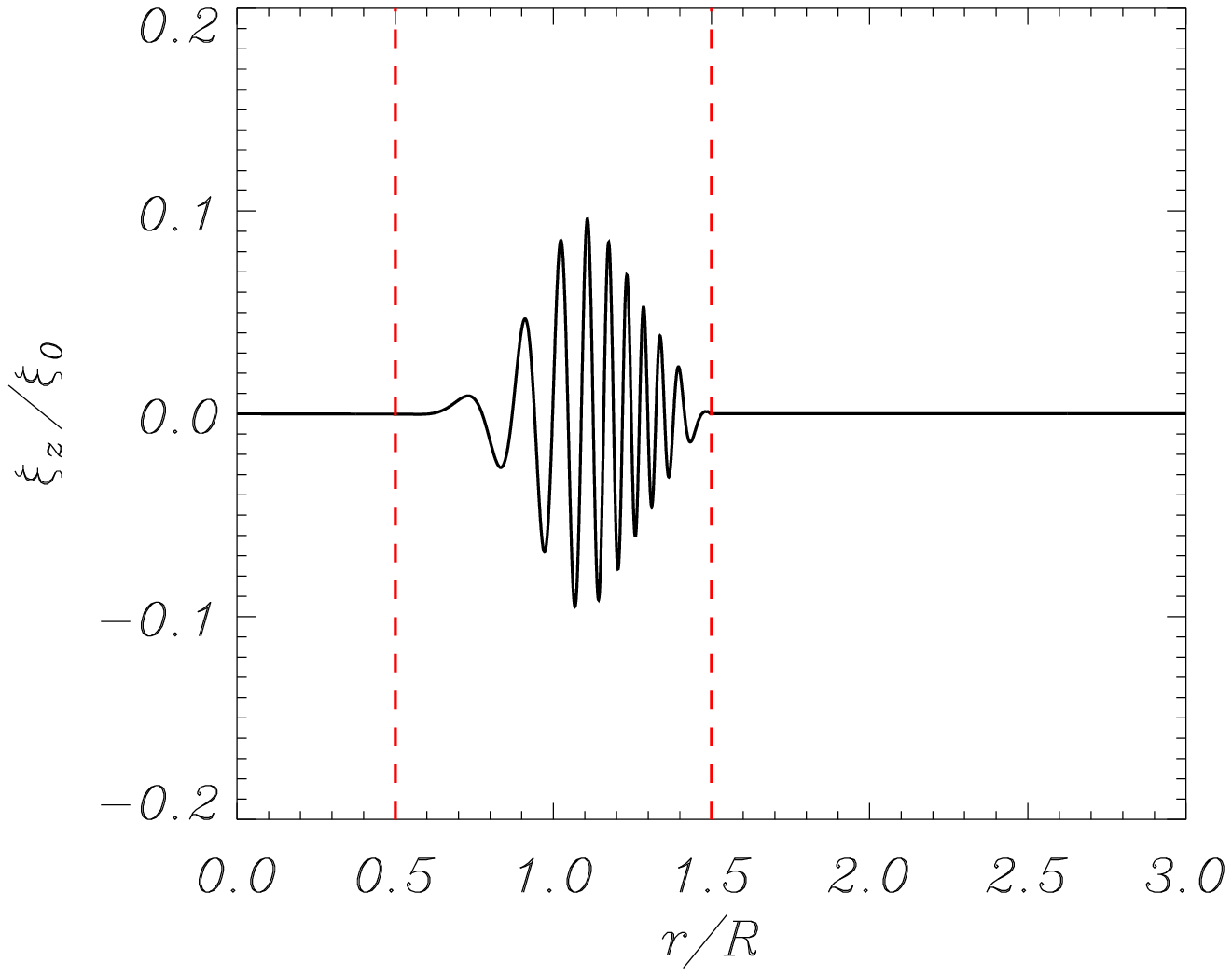}
	\caption{Same as Figure~\ref{fig:res1} but with $l/R=1$. Movie can be downloaded from \url{http://www.uib.es/depart/dfs/Solar/movies_SolerTerradas2015.zip}}
	\label{fig:res2}
\end{figure*}

The temporal evolution of the displacement components displayed in Figures~\ref{fig:res1} and \ref{fig:res2} shows that,  as time evolves, the amplitude of the oscillations in the nonuniform layer increases, while the amplitude in the internal and external plasmas decreases. The  global transverse oscillation damps and, after several periods, the oscillations are essentially polarized in the azimuthal direction. In other words, the initially transverse motion becomes rotational.  As time increases, the azimuthal component of the displacement in the nonuniform layer clearly dominates the dynamics, while both radial and longitudinal components are much smaller than the azimuthal component. This behavior is fully consistent with the presence of an energy flux towards the nonuniform layer due to the process of resonant absorption  \citep[see, e.g.,][]{goossens2013}.

\begin{figure*}
\centering
\includegraphics[width=1.95\columnwidth]{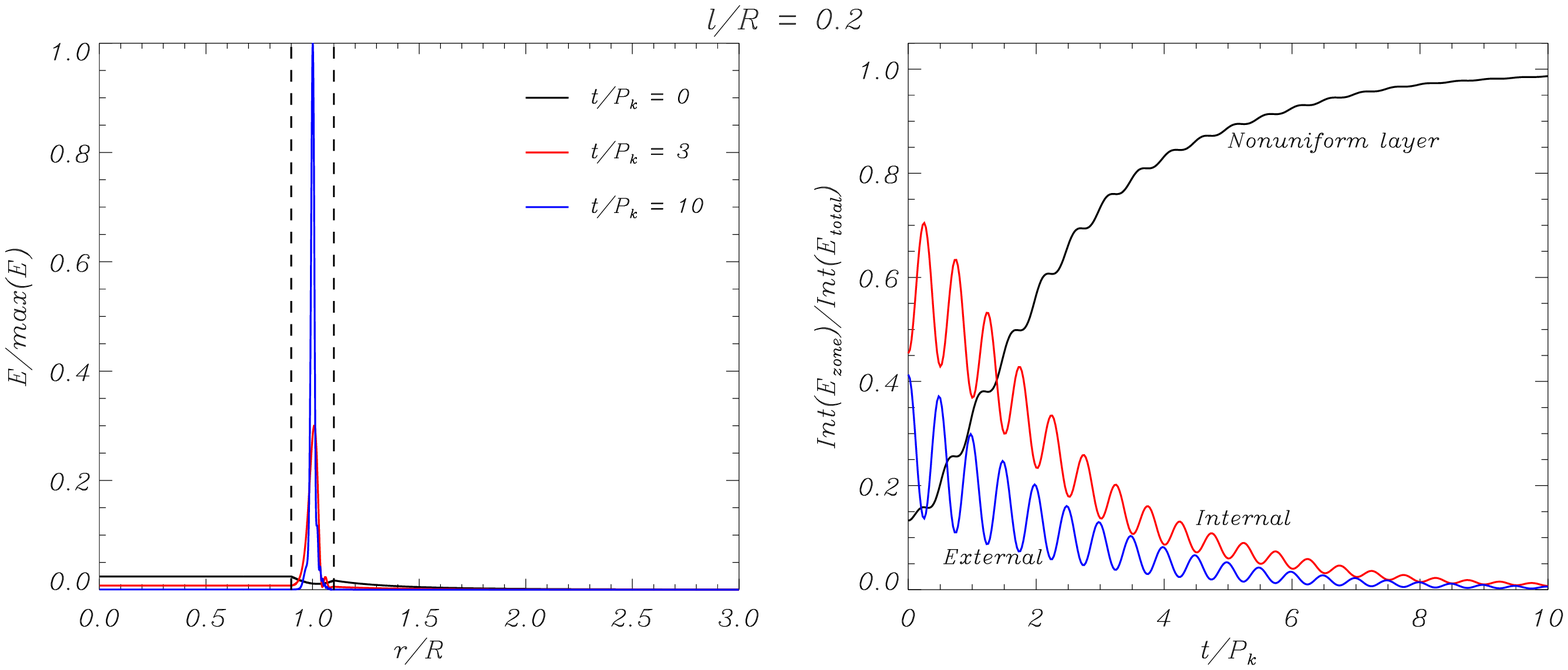}
\includegraphics[width=1.95\columnwidth]{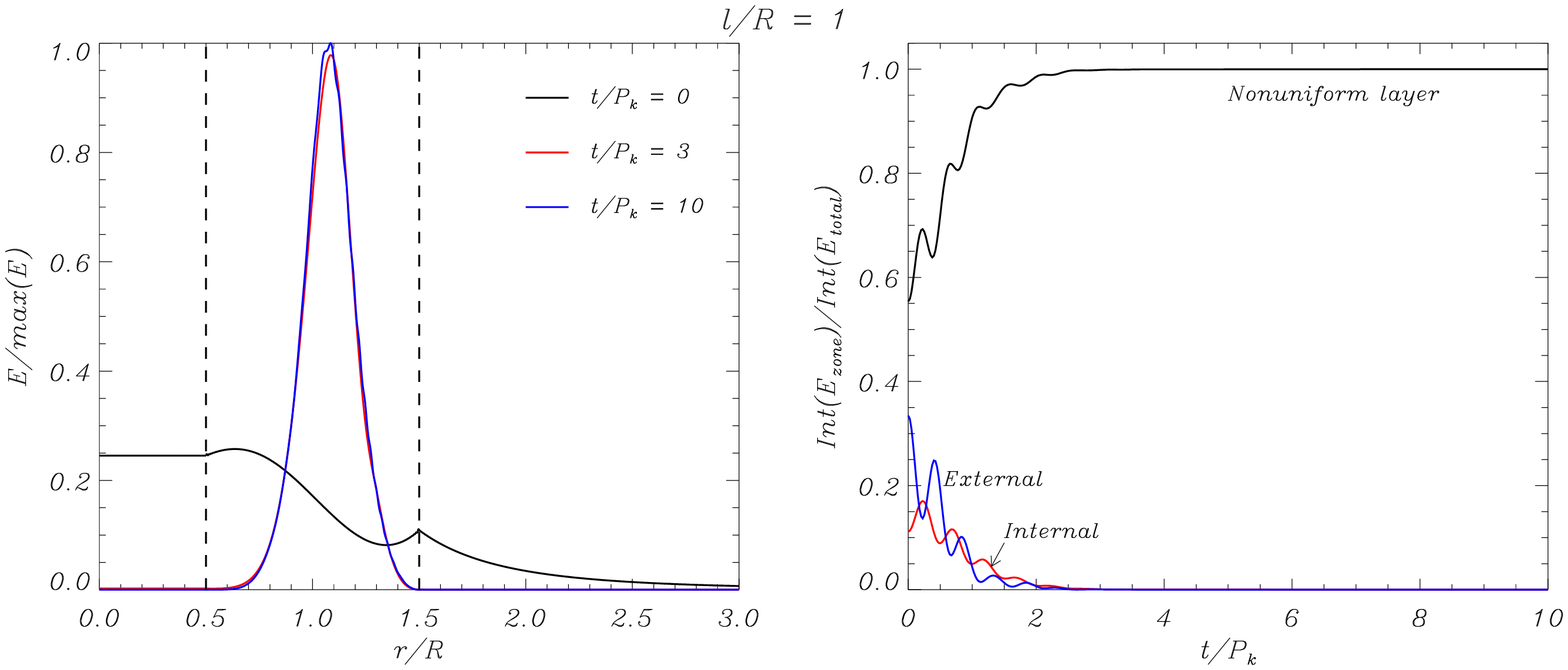}
	\caption{Energy density as function of position at three different times (left) and integrated energy in each region of the flux tube as a function of time (right) in a nonuniform tube with $l/R=0.2$ (top) and $l/R=1$ (bottom). The remaining parameters are the same as in Figure~\ref{fig:res1}. }
	\label{fig:energy}
\end{figure*}

To further illustrate the energy flux towards the nonuniform boundary of the tube due to resonant absorption, we compute the total (kinetic + magnetic) energy density, $E$, as \citep[see, e.g.,][]{walker2005}
\begin{equation}
E = \frac{1}{2} \left( \rho \left| {\bf v}' \right|^2 + \frac{1}{\mu}\left| {\bf B}' \right|^2 \right), \label{eq:energy}
\end{equation}
where $ {\bf v}' = \partial \xii / \partial t$ and ${\bf B'}=B \partial \xii / \partial z$ are the velocity and magnetic field perturbations, respectively. Figure~\ref{fig:energy} (left panels) shows the evolution with time of the spatial distribution of the energy density for the oscillations displayed in Figures~\ref{fig:res1} and \ref{fig:res2}. As time increases, more and more energy is fed into the nonuniform layer. Eventually, the shape of the energy density becomes a Gaussian-like function localized around the nonuniform boundary of the tube.  An equivalent  result can be seen in Figure~8 of \citet{terradas2006} obtained from numerical simulations, and also in Figure~7 of \citet{mann1995} for the case of magnetospheric waves.

 The right panels of Figure~\ref{fig:energy} display the integrated energy as function of time in each region of the flux tube, namely the internal plasma, the nonuniform layer, and the external plasma. The amount of energy localized in the internal and external plasmas becomes negligible at sufficiently large times. The efficiency of resonant absorption increases as  the nonuniform layer gets thicker \citep[see, e.g.,][]{rae1981,lee1986,goossens1992,rudermanroberts2002}. For this reason, the process of energy transfer is faster in the case of a thick layer  than in the case of a thin layer. The energy flux towards the boundary of the tube is very fast when the tube is largely inhomogeneous. For instance, for $l/R=1$ practically all the available energy is already localized around the boundary of the tube after  two periods of the kink oscillation. The process is slower for $l/R = 0.2$, but the final result is the same: all the wave energy is  transferred to the nonuniform boundary of the tube.

\subsection{Phase mixing and energy cascade to small scales}
\label{sec:phase}

\begin{figure*}
\centering
\includegraphics[width=.95\columnwidth]{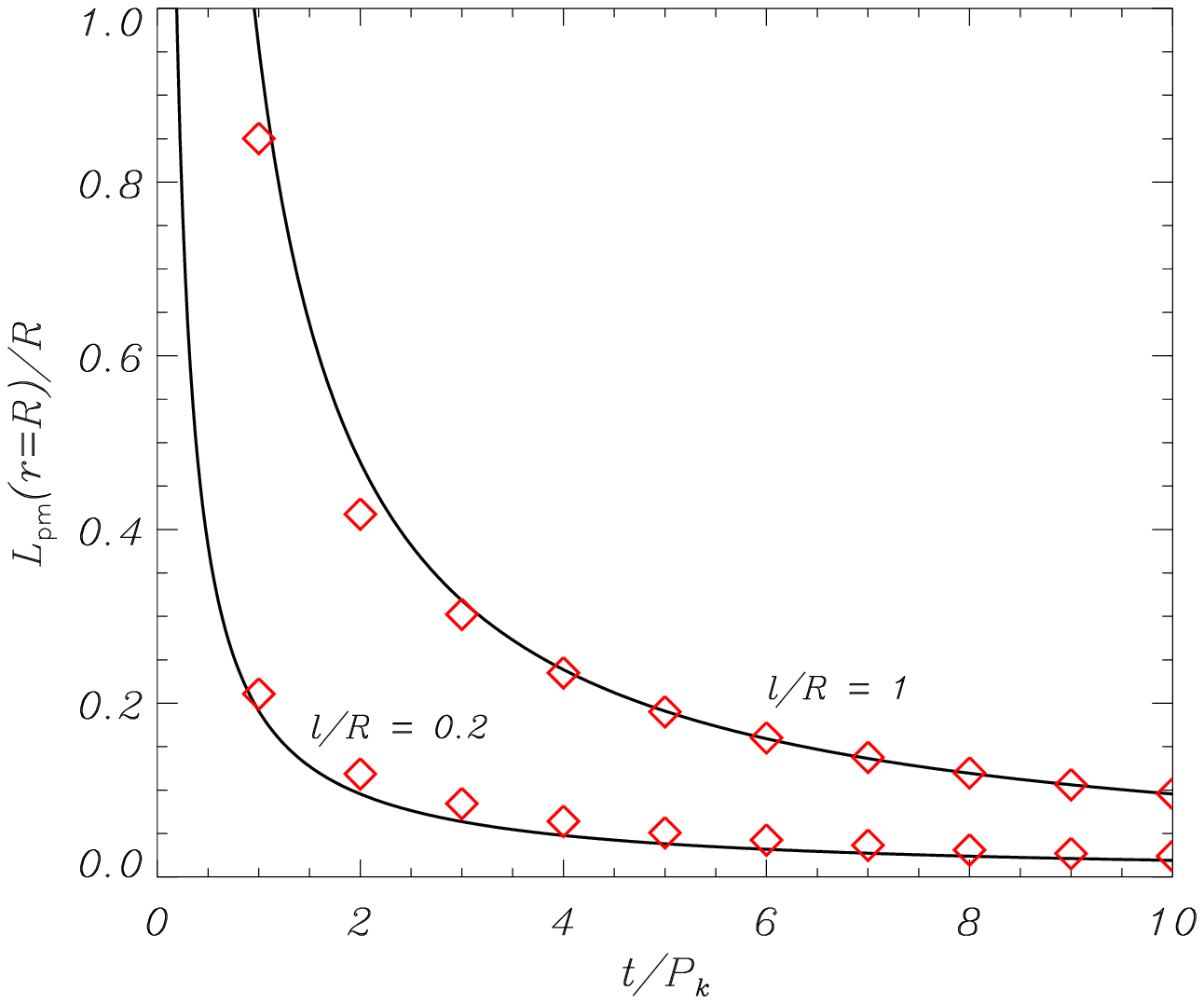}
\includegraphics[width=.95\columnwidth]{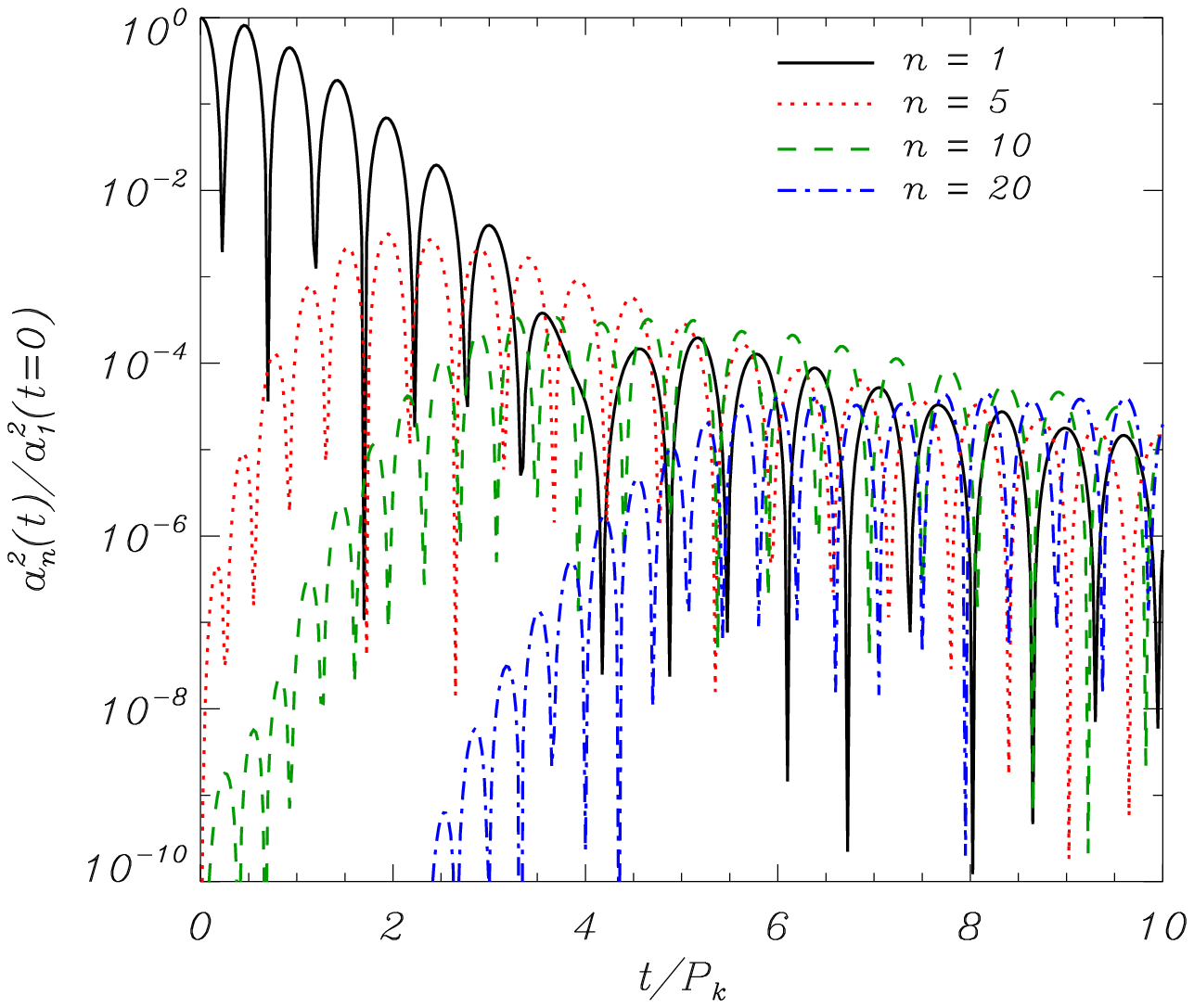}
	\caption{Left: Small spatial scale generated due to phase mixing at $r\approx R$. The solid line corresponds to Equation~(\ref{eq:lph}) and the symbols are the spatial scale directly estimated from the evolution of the displacement in the modal expansion method. Right: Energy cascade from large scales to small scales. Temporal evolution of $a_n^2$ (in logarithmic scale) for $n=$~1, 5, 10, and 20 in a flux tube with $l/R=1$. }
	\label{fig:cascade}
\end{figure*}

Small spatial scales are generated in the inhomogeneous layer as the process of resonant absorption  feeds energy into the nonuniform boundary of the tube. The temporal evolution of the displacement clearly shows how these small spatial scales evolve due to phase mixing. Smaller and smaller scales are generated as time increases \citep[see, e.g.,][]{pritchett1978,heyvaerts1983}. \citet{mann1995} derived an approximate expression for the length scale generated due to phase mixing, $L_{\rm pm}$, at a given position, $r=r_0$, and as a function of time. The expression is 
\begin{equation}
L_{\rm pm} = \frac{2\pi}{\left| \partial \omegaA/ \partial r  \right|_{r=r_0}\, t}, \label{eq:lph}
\end{equation}  
where $\left| \partial \omegaA/ \partial r  \right|_{r=r_0}$ is the absolute value of the  gradient of the Alfv\'en frequency at $r=r_0$. Figure~\ref{fig:cascade} (left panel) displays $L_{\rm pm}$ at $r=R$ as a function of time computed from Equation~(\ref{eq:lph}). In that graph, we overplot using symbols the length scale directly estimated from the evolution of the displacement obtained from the modal expansion method. To do that, we have taken the azimuthal component of the displacement and, for a fixed time, have measured the distance between the two consecutive extrema closer to $r=R$. We find a remarkable agreement between the length scale  estimated by this method and Equation~(\ref{eq:lph}). As expected, the phase-mixing length scale gets shorter as time increases. We also see that the thinner the nonuniform layer, the shorter  $L_{\rm pm}$. This last result is consistent with the fact that the density profile is steeper for a thin layer than for a thick layer. Hence, the thinner the nonuniform layer, the larger the gradient of the Alfv\'en frequency, and so the smaller the right-hand side of Equation~(\ref{eq:lph}) for a fixed time.

In the modal expansion, different spatial scales are associated to Fourier modes of different order. Long/short scales are associated to Fourier modes of low/high order. By substituting the Fourier expansion of the displacement (Equation~(\ref{eq:seriessturm})) into the expression for the energy density (Equation~(\ref{eq:energy})), it can be shown that the contribution of the $n$-th Fourier mode to the total energy is proportional to $a_n^2$. Figure~\ref{fig:cascade} (right panel) displays the temporal evolution of  $a_n^2$ for $n=$~1, 5, 10, and 20 in a flux tube with $l/R=1$.  This Figure evidences  the energy cascade from large scales to small scales due to phase mixing.  Fourier modes of high order (i.e., small scales) gain weight in the total energy as time increases, while Fourier modes of low order (i.e., large scales) lose weight. As time evolves, energy is distributed among smaller and smaller scales.

The process of phase mixing works indefinitely in our model since there is no dissipative mechanism included.  On purpose we used the ideal MHD equations to observe the process of energy cascade. Let us consider the scenario in which magnetic diffusion is included. We can relate $L_{\rm pm}$ with the typical length scale, $l_0$, that appears in the definition of the magnetic Reynolds number (Equation~(\ref{eq:reynolds})). Then, it can be seen that $R_m \sim t^{-1}$  owing to the effect of phase mixing, i.e., magnetic diffusion becomes more and more efficient as time increases.  In reality, the behavior of the MHD kink waves would be  essentially the same as in the ideal case studied here until the generated spatial scales due to phase mixing are sufficiently small for  diffusion to become efficient \citep[see][]{goedbloed2004}. Then, wave energy can be deposited in the background medium and so heat the plasma via Ohmic/viscous heating \citep[see, e.g.,][]{poedts1989,poedts1990,poedts1994,ofman1995}. An equivalent discussion is given by \citet{mann1995} for the case of dissipation in the magnetosphere due to Pedersen conductivity.

\section{DAMPING OF THE GLOBAL KINK MOTION AND COMPARISON TO THE QUASI-MODE}
\label{sec:quasi}
 
Here, we focus on the damping of the global transverse motion and analyze how the amplitude of the displacement changes with time near the axis of the flux tube. The purpose of this Section is to compare the results from the modal expansion method, in which there are no global modes, with those of the usual global quasi-mode approach.

The usual procedure to theoretically investigate the damping of  MHD kink waves in a transversely nonuniform flux tube is based on the assumption that there exists a global kink  mode. In other words, it is assumed that a nonuniform tube supports a global kink mode that is the  direct descendant of the kink mode of an equivalent uniform tube \citep[see, e.g.,][]{edwin1983}. This hypothetical global kink mode represents a coordinated transverse motion of the flux tube, so that the whole plasma is assumed to oscillate at the same frequency \citep[for more details see][and references therein]{tirry1996}.  The frequency of the  global kink mode is complex and the imaginary part of the  frequency is the damping  time scale due to resonant absorption \citep[see, e.g.,][]{sedlacek1971,tataronis1973a,rae1981,rae1982,goossens1992,rudermanroberts2002,paperI}.  However,  complex eigenvalues are not possible in ideal MHD \citep[see, e.g.,][]{poedts1991,goedbloed2004}. A complex frequency cannot possibly correspond to a true ideal normal mode of the flux tube. For this reason, the complex mode that physically represents a damped global motion of the flux tube is called a `quasi-mode' or a `virtual mode' \citep[see also][]{sedlacek1989}.

\begin{figure*}
\centering
\includegraphics[width=.95\columnwidth]{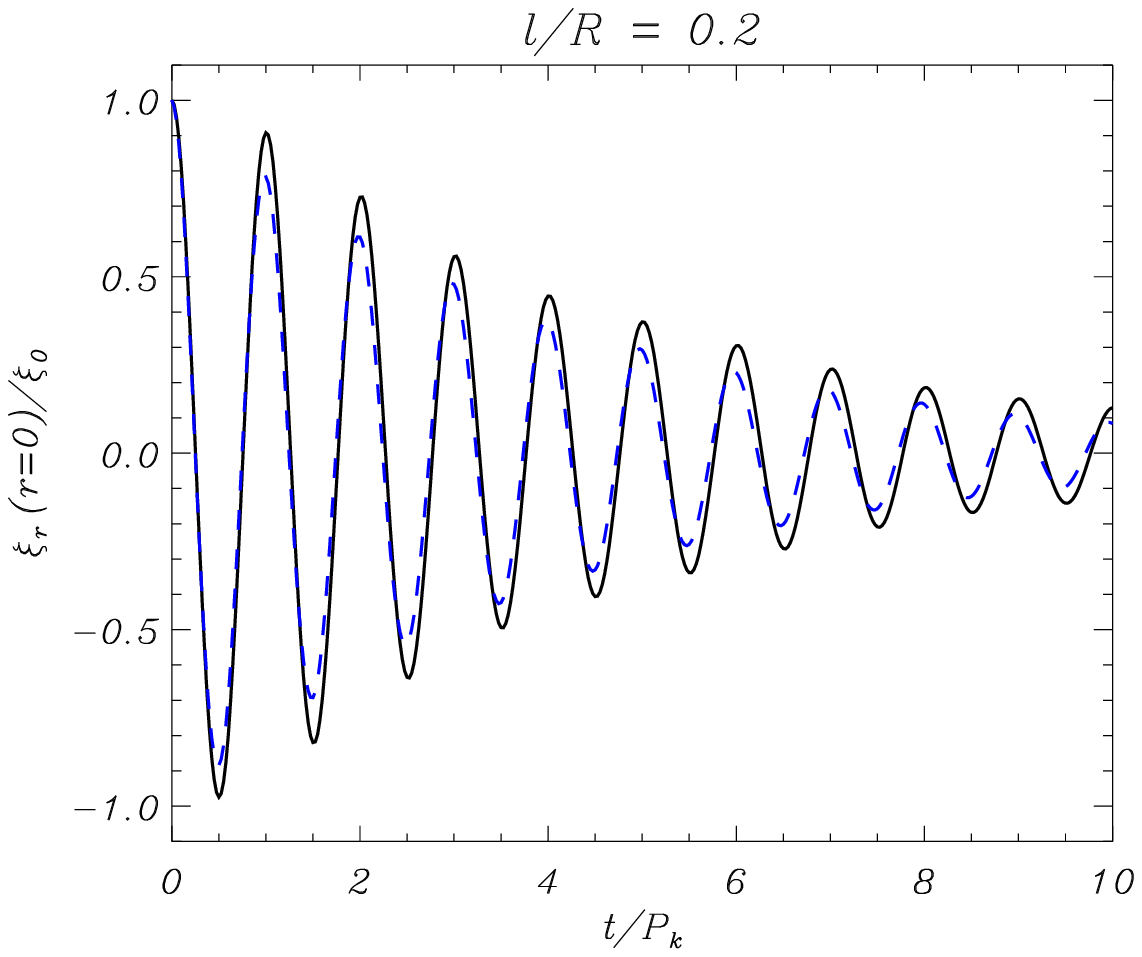}
\includegraphics[width=.95\columnwidth]{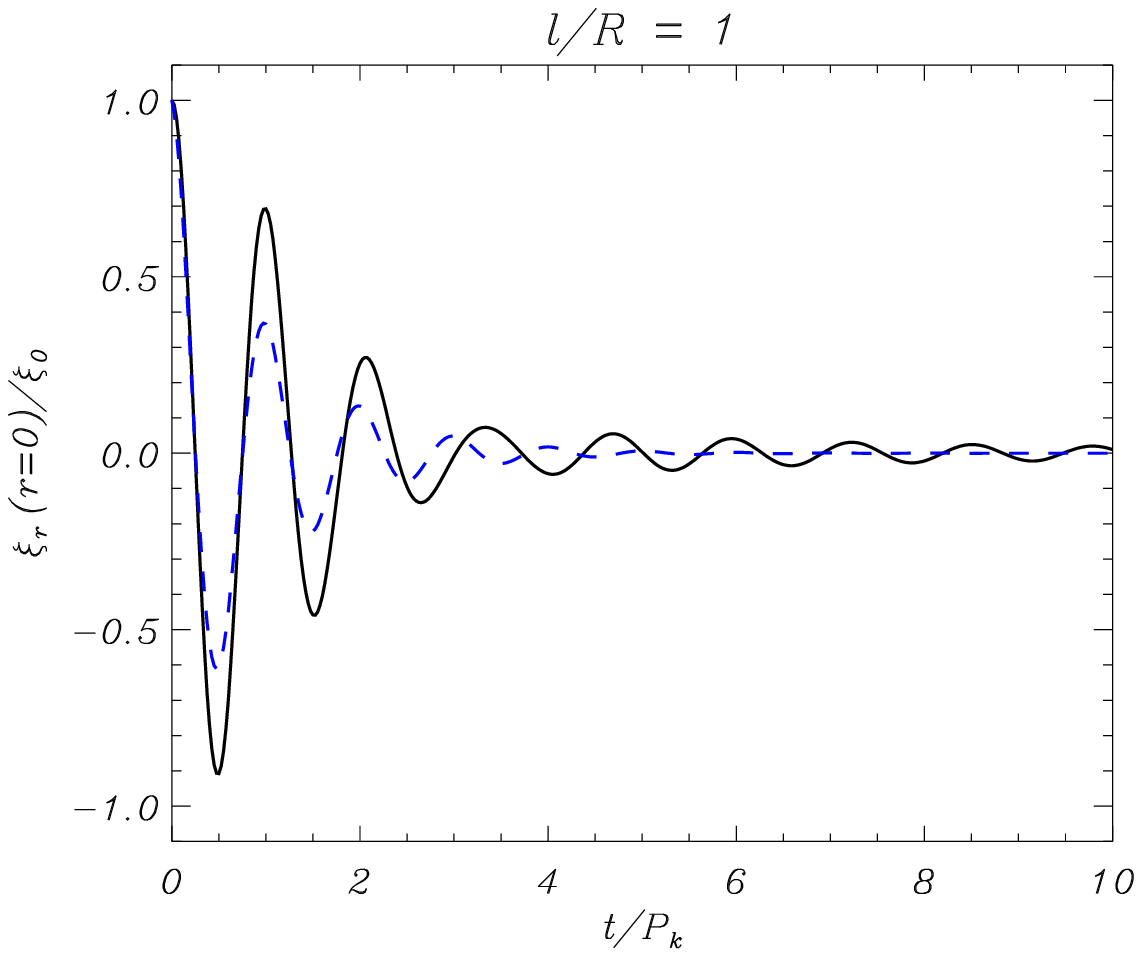}
	\caption{Temporal evolution of $\xi_r / \xi_0$ at $r=0$ in a nonuniform tube with $l/R = 0.2$ (left) and $l/R=1$ (right). The black solid line is the result from the modal expansion method and the blue dashed line is the dependence predicted by the quasi-mode (computed with the method of \citealt{paperI}). }
	\label{fig:ai}
\end{figure*}

\begin{figure*}
\centering
\includegraphics[width=.95\columnwidth]{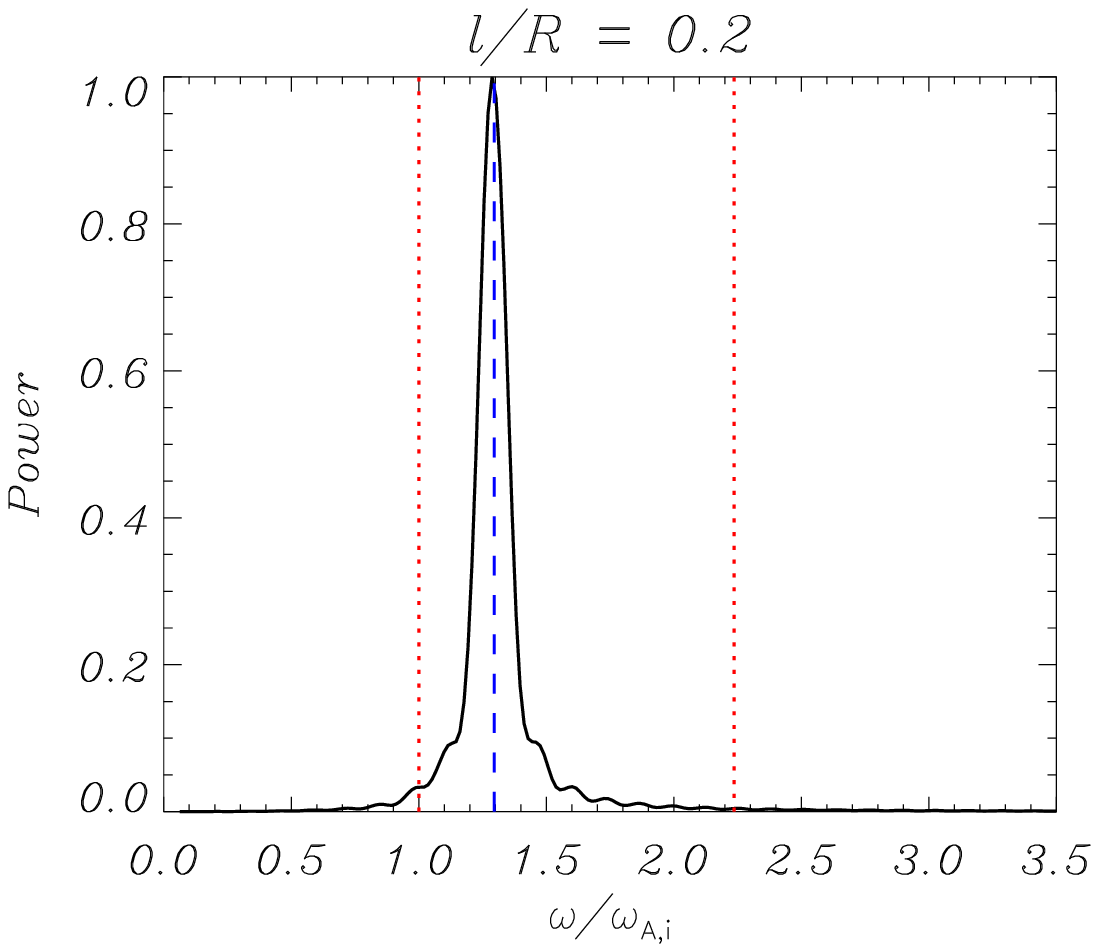}
\includegraphics[width=.95\columnwidth]{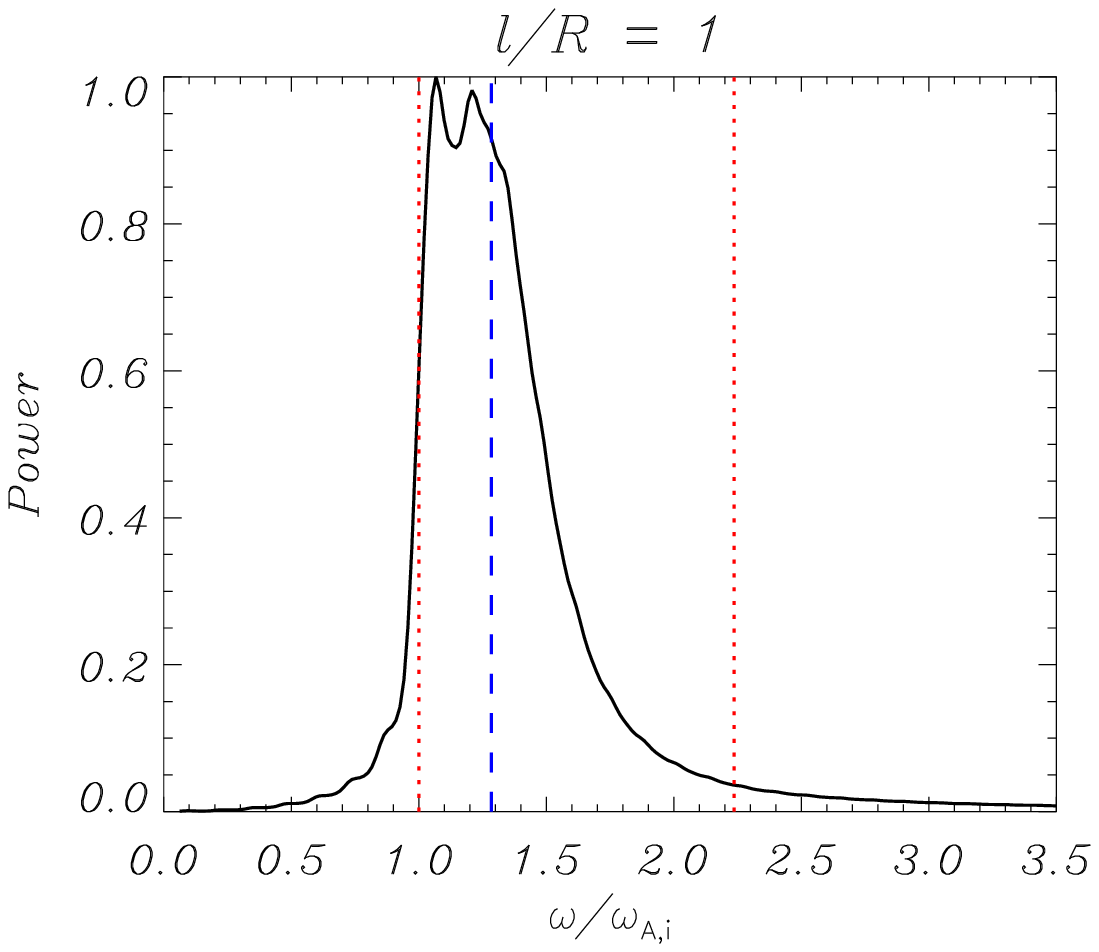}
	\caption{Power spectrum (normalized power vs. frequency) of $\xi_r$ at $r=0$  displayed in Figure~\ref{fig:ai}. The two vertical red dotted lines denote the internal and external Alfv\'en frequencies, while the vertical blue dashed line denotes the quasi-mode frequency (computed with the method of \citealt{paperI}).}
	\label{fig:period}
\end{figure*}

\begin{figure}
\centering
\includegraphics[width=.95\columnwidth]{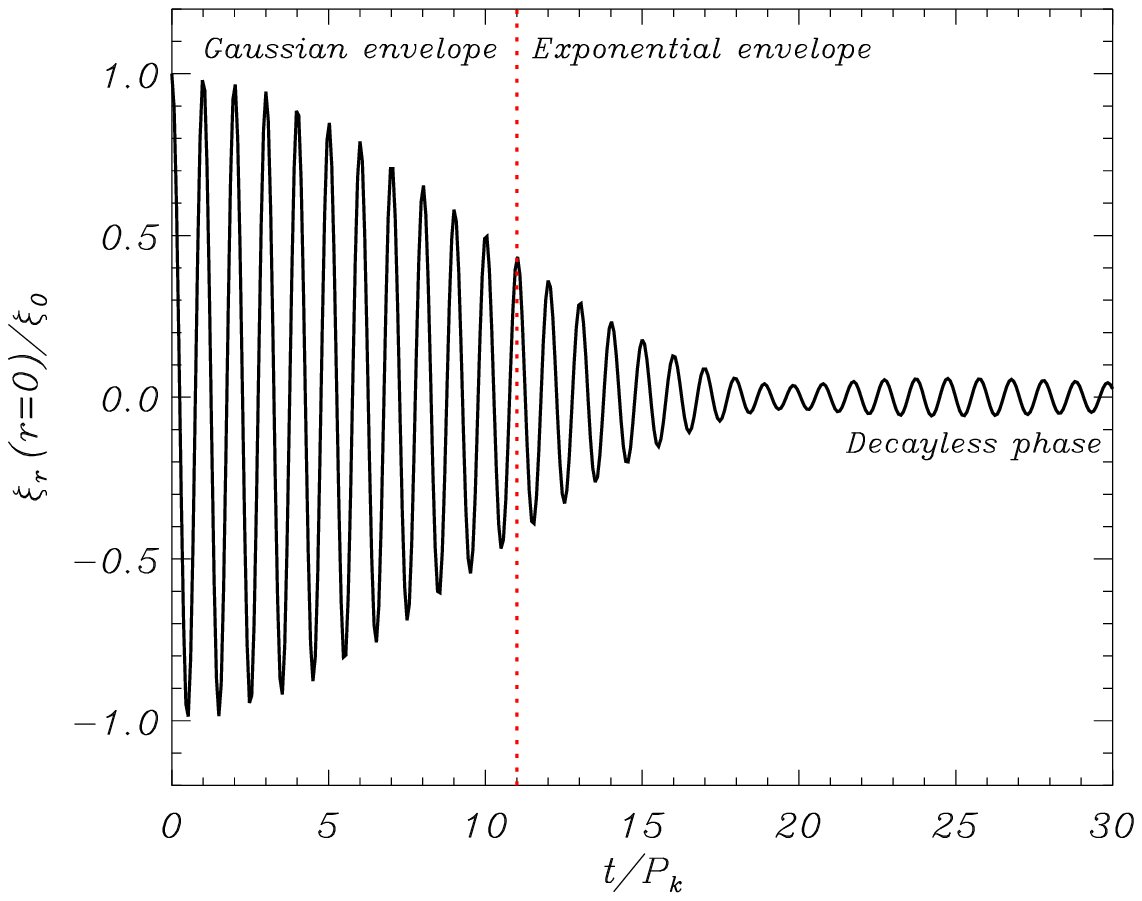}
	\caption{Temporal evolution of $\xi_r / \xi_0$ at $r=0$ in a nonuniform tube with $l/R = 1$ and a very low density contrast of $\rhoi/\rhoe = 1.2$. The vertical dotted line denotes the switch from Gaussian-like to exponential-like envelopes according to Equation~(10) of \citet{pascoe2013}. The decayless phase takes place for $t/P_k \gtrsim 20$.}
	\label{fig:ailow}
\end{figure}

 The quasi-mode temporal dependence is proportional to $\cos\left( \omega_{\rm qm} t \right) \exp\left( - \gamma t  \right)$, where $\omega_{\rm qm}$ and $\gamma$ are the quasi-mode frequency and damping rate, respectively. Here we compute the quasi-mode frequency and damping rate with the method described in \citet{paperI}. In short, we solve the singular  MHD equations in the nonuniform layer using the Method of Frobenius and find a `dispersion relation' for the quasi-mode \citep[see also][]{hollweg1990a}. The numerical solution to the quasi-mode `dispersion relation' provides us with $\omega_{\rm qm}$ and $\gamma$. For the sake of simplicity we omit the details of the quasi-mode computation and refer readers to \citet{paperI} for an extensive explanation of the procedure. Importantly, we note that the method of \citet{paperI} is general and uses no approximation to compute the quasi-mode and, therefore, it is applicable to thick nonuniform layers. 

\citet{andries2007} investigated the continuous spectrum of leaky MHD modes in a slab. These authors showed that the spectral measure  associated with the continuous  leaky spectrum is not a monotonic function of the frequency, but peaks appear in the spectral measure around specific frequencies (see their Figure~2). According to \citet{andries2007}, these specific frequencies correspond to the discrete damped leaky modes described in the literature \citep[e.g.,][]{cally1986}, while the characteristic time scale associated with the decay  of those continuum leaky modes is determined by the width of the peaks.  We can perform a clear parallelism between the results of \citet{andries2007} for leaky modes and those obtained here for resonantly damped quasi-modes. In Figure~\ref{fig:cs} we see that the peak in the frequency distribution of the  `Alfv\'en continuum modes' is wider for $l/R=1$ than for $l/R=0.2$. This result is consistent with the fact that the damping rate of the quasi-mode increases as the nonuniform layer gets thicker \citep[see, e.g.,][]{lee1986,rudermanroberts2002,paperI}, and is equivalent to the behaviour discussed by  \citet{andries2007} for the decay of leaky modes. We have overplotted in Figure~\ref{fig:cs} the quasi-mode frequency. When $l/R=0.2$ the quasi-mode frequency matches perfectly the sharp peak in the distribution of the  `Alfv\'en continuum modes'. However, when $l/R=1$ the quasi-mode frequency is smaller than the central frequency of the peak. When $l/R=1$  it is  unclear how the quasi-mode frequency is related to the distribution of the continuum frequencies. This result may suggest that the quasi-mode frequency is a good approximation to the effective frequency of the global oscillation when the transitional layer is thin, but it may become less representative when thick layers are considered.

In order to explore the accuracy of the quasi-mode compared to the actual temporal evolution, we display in Figure~\ref{fig:ai} the temporal evolution of $\xi_r$ at the axis of the flux tube. Again we consider two  values of the nonuniform layer thickness, namely a thin layer with $l/R = 0.2$ and a thick layer with $l/R=1$. These results correspond to the same  computations displayed in Figures~\ref{fig:res1} and \ref{fig:res2}, but now we focus on the behavior at $r=0$. For comparison, we overplot in Figure~\ref{fig:ai} the temporal dependence of the quasi-mode.  The modal expansion method and the quasi-mode  essentially provide the same result when the nonuniform layer is thin (see the left panel of Figure~\ref{fig:ai}). In that case the amplitude of $\xi_r$ near the flux tube axis decreases exponentially at the quasi-mode damping rate. However, as expected, the results are more different when the nonuniform layer is thick (see right panel of Figure~\ref{fig:ai}). The result from the modal expansion method shows two distinct phases. In the first phase, $\xi_r$ approximately oscillates at the quasi-mode frequency, but its amplitude does not decrease as fast as the quasi-mode predicts. In the second phase, $\xi_r$ oscillates at a frequency different from that of the quasi-mode and its amplitude remains roughly constant, i.e., the oscillations in this second phase are almost decayless. In the right panel of Figure~\ref{fig:ai} the switch from the first phase to the second phase takes place at $t/P_k \approx 3$. Hence, the quasi-mode offers a poor approximation to the actual transverse displacement of the flux tube axis in the case of thick nonuniform layers and large times.

To shed more light on the similarities and differences between the results from the  modal expansion method and those from the quasi-mode, we display in  Figure~\ref{fig:period}  the power spectra of the signals of Figure~\ref{fig:ai}. In the case of the result for a thin nonuniform layer (left panel), we find that the power spectrum has a pronounced peak around the quasi-mode frequency. This result confirms that the flux tube axis   oscillates with an effective frequency that corresponds to the quasi-mode frequency. However, contrary to the case for a thin layer, we do not find such a clear, sharp peak in the power spectrum for a thick layer (right panel). Although with some differences,  in both  cases the power spectrum at $r=0$ resembles the frequency distribution of the `Alfv\'en contiuuum modes' in the nonuniform layer (compare Figures~\ref{fig:cs} and \ref{fig:period}).

In  the power spectrum for thick layers (Figure~\ref{fig:period}, right panel), there is significant power around the quasi-mode frequency (as for thin layers) but we also find significant power in those frequencies near the internal Alfv\'en frequency. An explanation for the shape of the power spectrum  is found in the presence of the two phases of the oscillation discussed above (see again the right panel of Figure~\ref{fig:ai}). The scenario is as follows. In the first phase of the oscillation, the flux tube axis oscillates at an effective frequency that matches the quasi-mode frequency. The oscillation amplitude decreases rapidly  and then it starts the second phase with a roughly constant amplitude. The oscillation frequency in the second phase turns out to be very close to the internal Alfv\'en frequency.  Hence, both the quasi-mode frequency and the internal Alfv\'en frequency significantly contribute to the power spectrum when the nonuniform layer is thick. Conversely, no trace of the internal Alfv\'en frequency is found in the power spectrum when the nonuniform layer is thin since the quasi-mode dominates for the whole duration of the temporal evolution displayed in Figure~\ref{fig:ai} (left panel).

Finally, it is interesting to test whether the  modal expansion method is able to recover the initial Gaussian-like envelope of the displacement obtained in time-dependent numerical simulations \citep{pascoe2012,pascoe2013,hood2013,rudermanterradas2013}. The quasi-mode damping profile is purely exponential, but previous works showed that a Gaussian profile shows up in early stages of the oscillation if the density contrast is low. We show in Figure~\ref{fig:ailow} the temporal evolution of $\xi_r$ at $r=0$ for $l/R=1$ and $\rhoi/\rhoe = 1.2$. The initial Gaussian-shaped envelope  is evident. According to Equation~(10) of \citet{pascoe2013}, the switch from Gaussian to exponential envelopes occurs at $t/P_k \approx 11$ for the parameters considered in Figure~\ref{fig:ailow} and agrees well with the result displayed in that Figure. In addition, we again see in Figure~\ref{fig:ailow} the presence of the second, decayless phase of the oscillations when $t/P_k \gtrsim 20$.

\section{DISCUSSION}
\label{sec:dis}

In this paper, we have studied the temporal evolution of MHD kink waves in nonuniform magnetic flux tubes. Inspired by \citet{cally1991}, we have used a semi-analytic method based on expressing the Lagrangian displacement of the incompressible MHD kink wave as a superposition of discretized modes of the  Alfv\'en continuum. The behavior of the MHD kink wave obtained with this method matches that seen in full numerical simulations \citep[e.g.,][]{terradas2006,terradas2008,pascoe2013,goossens2014}. Hence, the present paper shows an alternative method that can be used in future works to include the full dynamics of MHD kink waves in, e.g., seismology applications or forward modelling.

The evolution of the MHD kink wave with time shows how the processes of resonant absorption and phase mixing operate. Due to the combined effect of both mechanisms, the initial global kink oscillation of the flux tube is eventually transformed into small-scale rotational motions in the nonuniform boundary of the tube. Resonant absorption transfers wave energy towards the nonuniform boundary, where phase mixing is responsible for the energy cascade from large spatial scales to small spatial scales. The two processes occur simultaneously and, indeed, they are intimately linked: both processes are caused by  plasma and/or magnetic field inhomogeneity across the flux tube. Such inhomogeneities naturally occur in magnetic flux tubes of the solar atmosphere.

Due to the use of the  linear ideal MHD equations, our analysis misses the very last step of the evolution of the MHD kink wave when the spatial scales generated by phase mixing are sufficiently small: heating of the background plasma due to Ohmic/viscous dissipation of wave energy. To study the heating, the  nonlinear dissipative MHD equations should be solved using, necessarily, full numerical simulations  \citep[see, e.g.,][]{poedts1989,poedts1990,poedts1994,ofman1995}. In this work we have studied the evolution of the MHD kink wave as an initial-value problem. Hence, the total energy available in the system is determined by the initial condition and, therefore, the maximum amount of energy that can be deposited in the plasma in the form of heat is limited. For instance, this scenario would correspond to post-flare kink oscillations of coronal loops \citep[e.g.,][]{nakariakov1999,aschwanden1999}. It is unlikely that such events play a significant role for the heating of the solar atmospheric plasma due to the limited amount of energy available. However, the results of this paper are also applicable to the case of propagating MHD kink waves continuously driven by footpoint motions \citep[see, e.g.,][]{tirry1997,degroof2002,pascoe2010,pascoe2012,pascoe2013,soler2011strat}. That scenario would be consistent with the ubiquitously observed waves propagating along coronal waveguides \citep[e.g.,][]{tomczyk2007,tomczyk2009}. The process of resonant absorption and phase mixing operate in the same way for standing and propagating waves but, in the case of periodically driven propagating waves, there is a continuous energy input and the dissipation of wave energy may provide a sustained plasma heating over time. This, however, should be checked by using self-consistent nonlinear simulations.

We have compared the actual temporal evolution obtained from the modal expansion method with the usual quasi-mode approach.  In summary, the quasi-mode offers only a partial view of the actual behavior of  MHD kink waves in nonuniform tubes. The quasi-mode  provides an effective frequency for the global oscillation of the flux tube and a time scale for which the  coordinated plasma motion loses its global character \citep[see][]{rae1981,lee1986}, but  the quasi-mode is unable to recover some important features of the full temporal evolution. Concerning the apparent damping of the global kink motion,  the quasi-mode is quite accurate when the nonuniform layer is thin, but it becomes less and less accurate as the thickness of the layer increases. This result may have implications for the accuracy of the seismology schemes for kink oscillations that are based on the quasi-mode, even when the schemes use no approximation and numerically compute the exact quasi-mode frequency and damping rate \citep[see][]{arregui2007seis,paperII}. Another limitation is that the quasi-mode does not describe the building up of small spatial scales and the phase mixing process in the nonuniform layer. Therefore, the quasi-mode provides no information about how the energy from the global  motion is redistributed among small spatial scales.  The quasi-mode becomes a true normal mode in dissipative MHD \citep[see, e.g.,][]{steinolfson1985,tirry1996} although for realistically small diffusivity the dissipative normal mode does not have the global character of the ideal quasi-mode and is indistinguishable from an ordinary resistive Alfv\'en mode  \citep{vandoorsselaere2007,paperI}. Having these issues in mind is crucial to correctly interpret the quasi-mode in physical terms and to reconcile ideal quasi-modes  with  dissipative eigenmodes \citep[see also the discussion in][]{goossens2014}.

In the case of largely nonuniform tubes, the temporal evolution of the transverse displacement of the tube axis reveals that, after a first phase dominated by the fast damping of the global oscillation, there is a second phase where the oscillations have a lower but roughly constant amplitude. This second phase might be related to the recently discovered decayless kink oscillations of coronal loops \citep[see][]{nistico2013,anfi2013}. Although that relation is speculative at present, it might be worth that future works investigate a possible link between those recent observations and the results discussed here.

\acknowledgements{
We thank Prof. Paul Cally  for reading a draft of this paper and for giving helpful comments. We acknowledge support from MINECO  through project AYA2011-22846, from CAIB through the `Grups Competitius' program, and from FEDER funds. R.S. also acknowledges support from MINECO through a `Juan de la Cierva' grant, from MECD through project CEF11-0012, and from the `Vicerectorat d'Investigaci\'o i Postgrau' of the UIB. J.T. also acknowledges support from the Spanish `Ministerio de Educaci\'on y Ciencia' through a `Ram\'on y Cajal' grant.}

\bibliographystyle{apj} 
\bibliography{refs}

\end{document}